\documentclass[twocolumn,tighten,times]{aastex62}
\usepackage{amsmath}
\usepackage{longtable}
\submitjournal{ApJ}
\shorttitle{Nitrogen Abundance Distribution in the inner Milky Way } \shortauthors{Pineda
  et al.}

\begin{document}
\title{Nitrogen Abundance Distribution in the Inner  Milky Way}

\author[0000-0001-8898-2800]{Jorge L. Pineda} 
\affiliation{Jet Propulsion Laboratory, California Institute of Technology, 4800 Oak Grove Drive, Pasadena, CA 91109-8099, USA} 
\author[0000-0002-8395-3557]{Shinji Horiuchi}
\affil{CSIRO Space  \& Astronomy/NASA Canberra Deep Space Communication Complex, PO Box 1035, Tuggeranong ACT 2901, Australia}
\author[0000-0002-7045-9277]{L. D. Anderson}
\affil{Department of Physics and Astronomy, West Virginia University, Morgantown, WV 26506, USA}
\affil{ Center for Gravitational Waves and Cosmology, West Virginia University, Chestnut Ridge Research Building, Morgantown, WV 26505, USA}
\affil{ Green Bank Observatory, P.O. Box 2, Green Bank, WV 24944, USA }
\author[0000-0001-8061-216X]{Matteo Luisi}
\affil{Department of Physics, Westminster College, New Wilmington, PA 16172, USA}
\affil{ Center for Gravitational Waves and Cosmology, West Virginia University, Chestnut Ridge Research Building, Morgantown, WV 26505, USA}
\author{William D. Langer}
\affil{Jet Propulsion Laboratory, California Institute of  Technology, 4800 Oak Grove Drive, Pasadena, CA 91109-8099, USA}
\author[0000-0002-6622-8396]{Paul F. Goldsmith}
\affil{Jet Propulsion Laboratory, California Institute of  Technology, 4800 Oak Grove Drive, Pasadena, CA 91109-8099, USA}
\author[0000-0003-1798-4918]{Thomas B. H. Kuiper}
\affil{Jet Propulsion Laboratory, California Institute of  Technology, 4800 Oak Grove Drive, Pasadena, CA 91109-8099, USA}
\author[0000-0003-2649-3707]{Christian Fischer}
\affiliation{Deutsches SOFIA Institut, Pfaffenwaldring 29, 70569 Stuttgart, Germany}
\author[0000-0002-3866-414X]{Yan Gong}
\affiliation{Max-Planck-Institut für Radioastronomie,Auf dem H\"ugel 69, 53121 Bonn, Germany}
\author{Andreas Brunthaler}
\affiliation{Max-Planck-Institut für Radioastronomie,Auf dem H\"ugel 69, 53121 Bonn, Germany}
\author[0009-0009-0025-9286]{Michael Rugel}
\altaffiliation{M.R.R. is a Jansky Fellow of the National Radio Astronomy Observatory, USA}. 
\affiliation{Center for Astrophysics, Harvard \& Smithsonian, 60 Garden St., Cambridge, MA 02138, USA}
\affiliation{National Radio Astronomy Observatory, 1003 Lopezville Rd, Socorro, NM 87801, USA}
\affiliation{Max-Planck-Institut für Radioastronomie,Auf dem H\"ugel 69, 53121 Bonn, Germany}
\author[0000-0001-6459-0669]{Karl M. Menten}
\affiliation{Max-Planck-Institut für Radioastronomie,Auf dem H\"ugel 69, 53121 Bonn, Germany}

\correspondingauthor{Jorge L. Pineda}
\email{Jorge.Pineda@jpl.nasa.gov}




\begin{abstract}
    We combine a new Galactic plane survey of Hydrogen Radio
    Recombination Lines (RRLs) with far--infrared (FIR) surveys of
    ionized Nitrogen, N$^{+}$, to determine Nitrogen abundance across
    Galactic radius.  RRLs were observed with NASA DSS--43 70m antenna
    and the Green Bank Telescope in 108 lines--of--sight spanning
    $-135$\degr$< l < 60$\degr, at $b=0$\degr.  These positions were
    also observed in [N\,{\sc ii}] 122$\mu$m and 205$\mu$m lines with
    the {\it Herschel Space Observatory}.  Combining RRL and [N\,{\sc
        ii}] 122$\mu$m and 205$\mu$m observations in 41 of 108 samples
    with high signal--to--noise ratio, we studied ionized Nitrogen
    abundance distribution across Galactocentric distances of
    0--8\,kpc.  Combined with existing Solar neighborhood and Outer
    galaxy N/H abundance determinations, we studied this quantity's
    distribution within the Milky Way's inner 17 kpc for the first
    time.  We found a Nitrogen abundance gradient extending from
    Galactocentric radii of 4--17\,kpc in the Galactic plane, while
    within 0--4\,kpc, the N/H distribution remained flat.
    The gradient observed at large Galactocentric distances supports
    inside--out galaxy growth with the additional steepening resulting
    from variable star formation efficiency and/or radial flows in the
    Galactic disk, while the inner 4\,kpc flattening, coinciding with
    the Galactic bar's onset, may be linked to radial flows induced by
    the bar potential.  Using SOFIA/FIFI--LS and {\it Herschel}/PACS,
    we observed the [N\,{\sc iii}] 57$\mu$m line to trace doubly
    ionized gas contribution in a sub--sample of sightlines. We found
    negligible N$^{++}$ contributions along these sightlines,
    suggesting mostly singly ionized Nitrogen originating from low
    ionization H\,{\sc ii} region outskirts.
\end{abstract}

\keywords{ISM: molecules --- ISM: structure}

\section{Introduction}
\label{sec:introduction}

The regulation of star formation in galaxies is a key driver of galaxy
evolution. As galaxies evolve, gas from the circum--galactic medium
accretes into their disks, cooling down and forming dense molecular
clouds where star formation takes place. As massive stars form, their
radiative and mechanical feedback ionizes and disperses their
surrounding gas slowing the gravitational collapse and star
formation. Stellar feedback can also result in large scale outflows of
gas that is transported back to the circum--galactic medium which,
depending on how the energy and momentum from stellar feedback couples
with the gas, can either accrete back into the disk of the galaxy,
restarting the whole process, or be expelled to the inter--galactic
medium. The interplay between accretion and outflows of gas into and
out of the disk of galaxies, molecular cloud formation, and the
effects of stellar feedback into the interstellar medium determines
the efficiency and rate at which gas is converted into stars in
galaxies. Therefore understanding these processes individually as well
as their interplay is key for understanding the regulation of star
formation and the evolution of galaxies.

The distribution of elemental abundances in the disk of galaxies
provides a fundamental observational constraint for models of the
formation and evolution of galaxies.  Elements such as C, N, and O,
are products of ``primary" and "secondary" processes in massive and
inter-mediate mass stars \citep{Johnson2019} and therefore their
abundances are related to the type of stars, star formation rate, and
star formation history at a given location, with each element having a
different enrichment timescale.  Metallicities are typically referred
in terms of the Oxygen abundance with respect to hydrogen,
O/H. However the abundance of elements such as Nitrogen and Carbon can
also provide important insights on the chemical evolution of
galaxies. Oxygen is mostly formed in massive stars and its production
timescale is short ($\sim$0.1\,Gyr;
e.g. \citealt{Maionio2019}). Carbon and Nitrogen however, can also be
produced in intermediate mass stars, and the timescales for their
production are much longer than those for Oxygen
(1--10\,Gyr). Therefore, Nitrogen and/or Carbon abundances, alongside
that of Oxygen, can provide important information on the star
formation history in galaxies.

The distribution of elements in the disk of galaxies provides
important constraints on the growth of galaxies as it is related to
the radial gas accretion profile in galaxies, which is an important
parameter on galaxy chemical evolution models
\citep{Larson1976,Matteucci1989,Boissier1999,Pezzulli2016}.  Several
studies of the distribution of elemental abundances in the disk of
galaxies have been conducted using optical lines
\citep{Sanchez2014,Perez-Montero2016,Sanchez-Menguiano2016,Belfiore2017}.
\citet{Belfiore2017} studied a large sample of galaxies, finding that
the abundance of Oxygen relative to that of Hydrogen, O/H, decreases
with galactocentric distance with a slope that steepens with the
galaxy's stellar mass. For galaxies at the high end of the stellar
mass range in their sample ($M>10^{10.5}{\rm M}_\odot$), a flattening
of the O/H distribution is observed. They also studied the Nitrogen
abundance relative to that of Oxygen and found that the N/O ratio
increases with galactocentric distance and does not flatten in the
inner parts of massive galaxies.  These observations are interpreted
in the context of inside--out growth of galaxy disks in which the
central parts of massive galaxies reach a metallicity equilibrium, in
which metal production is balanced by metal consumption by star
formation expulsion by outflows, and while the outer part continues
accreting less enriched gas. Note however that, in the central parts
of massive galaxies, the dust extinction becomes significant and
therefore enriched gas close to the galaxy's center might be
unaccounted for by optical and near--IR observations
\citep{Puglisi2017,Calabro2018}.

It is important to test whether the properties of abundance
distributions observed in nearby galaxies also apply to the Milky Way.
Detailed studies of the properties of the interstellar medium and star
formation at high spatial resolution are currently only possible in
the Milky Way. Therefore, by studying the elemental abundance
distribution in the Milky Way, we can obtain a deeper insight into the
nature of these distributions, which in turn can be used to interpret
the results obtained over large samples of unresolved, external
galaxies.  The abundance distributions in the Milky Way have been
traditionally obtained observing H\,{\sc ii} regions with optical
lines in a variety of environments.  These observations which are
mostly focused on nearby H\,{\sc ii} regions, where dust extinction
obscures these optical lines moderately, have been used to infer that
the abundance of Oxygen and Nitrogen increase with Galactocentric
distance from the outer galaxy inward to $R_{\rm gal}= 4$\,kpc
\citep[][see also \citealt{Romano2022} and references therein]{Esteban2018, Arellano-Cordova2021}. Due to increased dust
extinction, optical studies are however unable to probe the inner
Galaxy, where most of the star formation takes place, and which is
thought to have formed during the early stages of the Milky Way's
evolution. Therefore, these observations have been unable to test the
flattening in the O/H distribution and a possible increase of the N/O
ratio observed in the central parts of external galaxies with similar
stellar masses as the Milky Way's.

\begin{figure*}[t]
\centering
\includegraphics[angle=0,width=0.95\textwidth,angle=0]{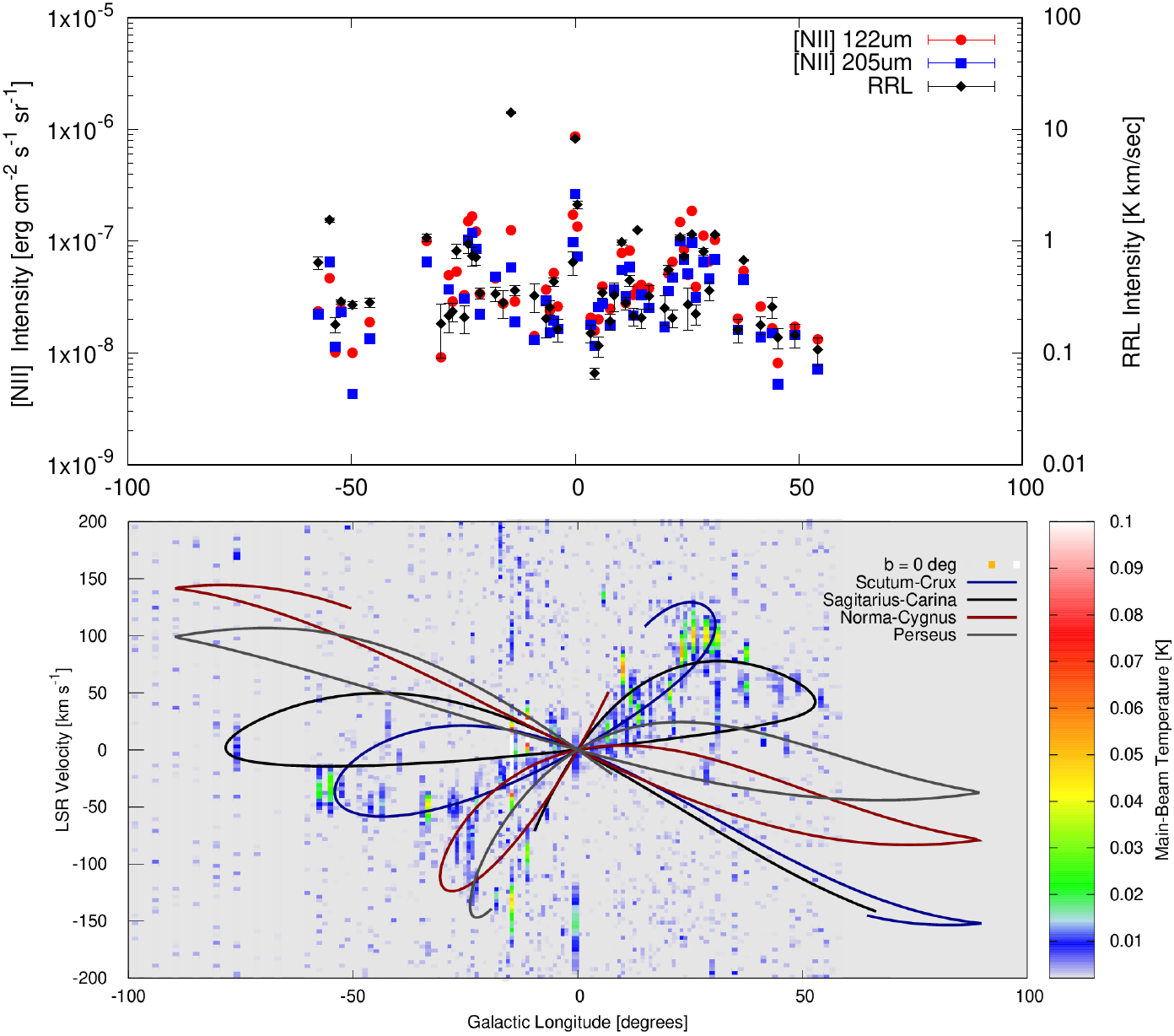}
\caption{({\it upper panel}) Distribution of the [N\,{\sc ii}]
  122$\mu$m, 205$\mu$m, and RRL emission as a function of Galactic
  longitude.  ({\it lower panel}) Position velocity map of Hydrogen
  recombination line emission for the observed GOT C+ LOS between $l =
  -$100\degr\ and 100\degr\ at b = 0\degr.  The position--velocity map
  is overlaid with projections of the Scutum--Crux,
  Sagittarius--Carina, Perseus, and Norma--Cygnus Milky Way spiral
  arms.}
\label{fig:pv_map}
\end{figure*}

Longer wavelength observations, including those of far--infrared fine
structure lines, radio continuum, and hydrogen radio recombination
lines can be combined to determine the abundance of elements with the
advantage that they are unobscured by dust, enabling elemental
abundance determination deep into the inner galaxy
\citep{Simpson1995,Afflerbach1997,Rudolph1997,Simpson2004,Rudolph2006,Balser2011}.
These observations however are limited to small samples mostly focused
in dense (ultra--)compact H\,{\sc ii} regions that might have complex
ionization structures, and thus require a large number of spectral
line observations to determine elemental
abundances. \citet{Goldsmith2015} presented a survey of 149
lines--of--sight observed uniformly in the Galactic plane in the
[N\,{\sc ii}] 205$\mu$m and 122$\mu$m lines at $b=0$\degr\ using the
{\it Herschel}/PACS instrument.  The [N\,{\sc ii}] 122$\mu$m/205$\mu$m
ratio provides an accurate determination of the electron density that
is independent of Nitrogen abundance and has a weak dependence on
electron temperature. The electron densities derived in their sample
show little scatter with a mean value of about 30\,cm$^{-3}$,
suggesting that these observations are tracing an extended moderate
density ionized gas component in the ISM which is likely to have
moderate far--UV (FUV) and extreme--UV (EUV) fields, and thus the
ionization structure is simpler than in compact H\,{\sc ii}
regions. In \citet{Pineda2019} we derived Nitrogen abundances in a
sample of 10 sight lines taken from the \citet{Goldsmith2015} survey
by combining [N\,{\sc ii}] and RRL observations. We found that the
distribution of Nitrogen abundances in the inner Galaxy derived from
our data has a linear slope that is consistent with that found in the
outer Galaxy in optical studies.  In this paper we present an analysis
of a larger sample to confirm these results and investigate the
distribution of Nitrogen in the inner Galactic plane.

We present in this paper a Galactic plane survey of Hydrogen
Recombination Lines, covering the range between $-135$\degr$< l <
60$\degr\ and $b=0$\degr\ using the NASA Deep Space Network Station 43
(DSS--43) in Canberra, Australia, and the Green Bank telescope.  The
lines--of--sight coincide with those observed by {\it Herschel}/HIFI
in [C\,{\sc ii}] \citep{Langer2010,Pineda2013} and by {\it
  Herschel}/PACS in [N\,{\sc ii}] 122$\mu$m and 205$\mu$m
\citep{Goldsmith2015}.  The high--velocity resolution observations of
RRLs will allow us to disentangle the source of [N\,{\sc ii}] emission
in the Galactic plane.  In this paper we focus on combining the RRL
data set with that observed in [N\,{\sc ii}] with {\it Herschel} to
derive the abundance of ionized Nitrogen with respect to ionized
Hydrogen in the inner parts of the Milky Way. This data set will be
also combined with SOFIA/FIFI--LS and {\it Herschel} observations of
the [N\,{\sc iii}] 57$\mu$m to account for the contribution of higher
ionization states of Nitrogen, such as N$^{++}$.

\begin{deluxetable*}{lcccccccc} 
\tabletypesize{\footnotesize} \centering \tablecolumns{9} \small
\tablewidth{0pt} \tablecaption{[N\,{\sc ii}], [N\,{\sc iii}], RRL, and radio continuum intensities for $l\geq0$\degr. } \tablenum{1}
\tablehead{
  \colhead{LOS} &
  \colhead{$l$}&
  \colhead{$b$} &
  \colhead{[N\,{\sc ii}] 122$\mu$m}&
  \colhead{[N\,{\sc ii}] 205$\mu$m}&
  \colhead{[N\,{\sc iii}] 57$\mu$m}&
  \colhead{RRL}&
  \colhead{RRL I.D.}&
  \colhead{$<$T$_{\rm C}>$}\\
  \colhead{} &
  \colhead{}&
  \colhead{} &
  \colhead{[$\times10^{-8}$]}&
  \colhead{[$\times10^{-8}$]}&
  \colhead{[$\times10^{-7}$]}&
  \colhead{}&
  \colhead{}&
  \colhead{free--free at 8.5GHz} \\
  \colhead{} &
  \colhead{[\degr]}&
  \colhead{[\degr]} &
  \colhead{[W\,m$^{-2}$\,sr$^{-1}$]}&
  \colhead{[W\,m$^{-2}$\,sr$^{-1}$]}&
  \colhead{[W\,m$^{-2}$\,sr$^{-1}$]}&
  \colhead{[K\,km\,s$^{-1}$]}&
  \colhead{}&
  \colhead{[K]}
  }
\startdata
G000.0+0.0 & 0.000 & 0.0 & 86.165 $\pm$ 0.097 & 26.426 $\pm$ 0.089 & 4.61 $\pm$ 0.76 & 8.23 $\pm$ 0.08 & H89$\alpha$ & 6.97 $\pm$ 0.01 \\
G000.5+0.0 & 0.500 & 0.0 & 13.509 $\pm$ 0.025 & 7.296 $\pm$ 0.068 & -- & 2.12 $\pm$ 0.16 &H89$\alpha$ & 2.17 $\pm$ 0.01 \\
G003.5+0.0 & 3.478 & 0.0 & 2.064 $\pm$ 0.005 & 1.775 $\pm$ 0.019 & -- & 0.15 $\pm$ 0.03 &H89$\alpha$ & 0.06 $\pm$ 0.01 \\
G004.3+0.0 & 4.348 & 0.0 & 1.580 $\pm$ 0.003 & 1.149 $\pm$ 0.021 & -- & 0.09 $\pm$ 0.04 &H89$\alpha$ & 0.05 $\pm$ 0.01 \\
G005.2+0.0 & 5.217 & 0.0 & 1.999 $\pm$ 0.012 & 2.566 $\pm$ 0.063 & -- & 0.12 $\pm$ 0.02 &H89$\alpha$ & 0.07 $\pm$ 0.01 \\
G006.1+0.0 & 6.087 & 0.0 & 3.927 $\pm$ 0.006 & 2.749 $\pm$ 0.014 & -- & 0.35 $\pm$ 0.04 &H89$\alpha$ & 0.19 $\pm$ 0.01 \\
G007.8+0.0 & 7.826 & 0.0 & 2.486 $\pm$ 0.004 & 1.780 $\pm$ 0.018 & -- & 0.19 $\pm$ 0.03 &H89$\alpha$ & 0.15 $\pm$ 0.01 \\
G008.7+0.0 & 8.696 & 0.0 & 3.689 $\pm$ 0.012 & 3.707 $\pm$ 0.083 & -- & 0.33 $\pm$ 0.10 &H89$\alpha$ & 0.34 $\pm$ 0.02 \\
G010.4+0.0 & 10.435 & 0.0 & 7.827 $\pm$ 0.017 & 5.505 $\pm$ 0.072 & 3.61 $\pm$ 0.40 & 0.98 $\pm$ 0.06 & H89$\alpha$ & 0.43 $\pm$ 0.02 \\
G011.3+0.0 & 11.304 & 0.0 & 2.798 $\pm$ 0.013 & 3.232 $\pm$ 0.097 & -- & 0.28 $\pm$ 0.04 &H89$\alpha$ & 0.09 $\pm$ 0.01 \\
G012.2+0.0 & 12.174 & 0.0 & 8.228 $\pm$ 0.017 & 5.864 $\pm$ 0.067 & $<$2.62 $\pm$ 0.87 & 0.45 $\pm$ 0.05 & H89$\alpha$ & 0.18 $\pm$ 0.01 \\
G013.0+0.0 & 13.043 & 0.0 & 3.306 $\pm$ 0.006 & 2.161 $\pm$ 0.019 & -- & 0.35 $\pm$ 0.05 &H89$\alpha$ & 0.20 $\pm$ 0.01 \\
G013.9+0.0 & 13.913 & 0.0 & 4.981 $\pm$ 0.013 & 3.793 $\pm$ 0.078 & -- & 1.26 $\pm$ 0.02 &H89$\alpha$ & 0.29 $\pm$ 0.02 \\
G014.8+0.0 & 14.783 & 0.0 & 4.055 $\pm$ 0.012 & 3.300 $\pm$ 0.059 & -- & 0.21 $\pm$ 0.04 &H89$\alpha$ & 0.16 $\pm$ 0.01 \\
G016.5+0.0 & 16.522 & 0.0 & 3.791 $\pm$ 0.005 & 2.537 $\pm$ 0.009 & -- & 0.32 $\pm$ 0.08 &H89$\alpha$ & 0.13 $\pm$ 0.01 \\
G020.0+0.0 & 20.000 & 0.0 & 1.715 $\pm$ 0.004 & 1.717 $\pm$ 0.011 & -- & 0.25 $\pm$ 0.07 &H89$\alpha$ & 0.12 $\pm$ 0.01 \\
G020.9+0.0 & 20.870 & 0.0 & 5.122 $\pm$ 0.012 & 3.570 $\pm$ 0.060 & -- & 0.56 $\pm$ 0.05 &H89$\alpha$ & 0.25 $\pm$ 0.01 \\
G021.7+0.0 & 21.739 & 0.0 & 6.533 $\pm$ 0.015 & 4.723 $\pm$ 0.064 & -- & 0.22 $\pm$ 0.04 &H89$\alpha$ & 0.22 $\pm$ 0.01 \\
G023.5+0.0 & 23.478 & 0.0 & 14.796 $\pm$ 0.016 & 9.960 $\pm$ 0.061 & $<$2.42 $\pm$ 0.81 & 1.08 $\pm$ 0.06 & H89$\alpha$ & 0.46 $\pm$ 0.01 \\
G024.3+0.0 & 24.348 & 0.0 & 8.428 $\pm$ 0.019 & 6.898 $\pm$ 0.063 & 3.67 $\pm$ 0.50 & 0.73 $\pm$ 0.03 & H89$\alpha$ & 0.30 $\pm$ 0.01 \\
G025.2+0.0 & 25.217 & 0.0 & 4.991 $\pm$ 0.014 & 5.093 $\pm$ 0.061 & -- & 0.27 $\pm$ 0.11 &H89$\alpha$ & 0.19 $\pm$ 0.01 \\
G026.1+0.0 & 26.087 & 0.0 & 18.625 $\pm$ 0.021 & 9.721 $\pm$ 0.075 & $<$4.64 $\pm$ 1.55 & 1.15 $\pm$ 0.03 & H89$\alpha$ & 0.45 $\pm$ 0.01 \\
G027.0+0.0 & 26.956 & 0.0 & 3.898 $\pm$ 0.006 & 3.141 $\pm$ 0.020 & -- & 0.22 $\pm$ 0.05 &H89$\alpha$ & 0.17 $\pm$ 0.01 \\
G028.7+0.0 & 28.696 & 0.0 & 11.173 $\pm$ 0.013 & 6.507 $\pm$ 0.065 & 2.64 $\pm$ 0.28 & 0.81 $\pm$ 0.06 & H89$\alpha$ & 0.52 $\pm$ 0.02 \\
G030.0+0.0 & 30.000 & 0.0 & 6.596 $\pm$ 0.013 & 4.577 $\pm$ 0.075 & -- & 0.38 $\pm$ 0.07 &H89$\alpha$ & 0.40 $\pm$ 0.02 \\
G031.3+0.0 & 31.277 & 0.0 & 10.271 $\pm$ 0.015 & 6.857 $\pm$ 0.068 & -- & 1.14 $\pm$ 0.02 &H89$\alpha$ & 0.45 $\pm$ 0.01 \\
G036.4+0.0 & 36.383 & 0.0 & 2.020 $\pm$ 0.011 & 1.602 $\pm$ 0.075 & -- & 0.16 $\pm$ 0.04 &H89$\alpha$ & 0.09 $\pm$ 0.01 \\
G037.7+0.0 & 37.660 & 0.0 & 5.400 $\pm$ 0.010 & 4.505 $\pm$ 0.073 & -- & 0.67 $\pm$ 0.01 &H89$\alpha$ & 0.25 $\pm$ 0.01 \\
G041.5+0.0 & 41.489 & 0.0 & 2.600 $\pm$ 0.003 & 1.387 $\pm$ 0.010 & -- & 0.19 $\pm$ 0.03 &H89$\alpha$ & 0.15 $\pm$ 0.01 \\
G044.0+0.0 & 44.043 & 0.0 & 1.658 $\pm$ 0.002 & 1.501 $\pm$ 0.020 & -- & 0.26 $\pm$ 0.06 &H89$\alpha$ & 0.08 $\pm$ 0.01 \\
G045.3+0.0 & 45.319 & 0.0 & 0.812 $\pm$ 0.002 & 0.525 $\pm$ 0.012 & -- & 0.14 $\pm$ 0.03 &H89$\alpha$ & 0.07 $\pm$ 0.01 \\
G049.1+0.0 & 49.149 & 0.0 & 1.716 $\pm$ 0.012 & 1.447 $\pm$ 0.055 & -- & 0.14 $\pm$ 0.03 &H89$\alpha$ & 0.15 $\pm$ 0.01 \\
G054.3+0.0 & 54.255 & 0.0 & 1.322 $\pm$ 0.003 & 0.716 $\pm$ 0.011 & -- & 0.11 $\pm$ 0.03 &H89$\alpha$ & 0.09 $\pm$ 0.01 
\enddata
\tablenotetext{}{}
\label{tab:intensities}
\end{deluxetable*}

\begin{deluxetable*}{lcccccccc} 
\tabletypesize{\footnotesize} \centering \tablecolumns{9} \small
\tablewidth{0pt} \tablecaption{[N\,{\sc ii}], [N\,{\sc iii}], RRL, and radio continuum intensities  for $l<0$\degr.} \tablenum{2}
\tablehead{
  \colhead{LOS} &
  \colhead{$l$}&
  \colhead{$b$} &
  \colhead{[N\,{\sc ii}] 122$\mu$m}&
  \colhead{[N\,{\sc ii}] 205$\mu$m}&
  \colhead{[N\,{\sc iii}] 57$\mu$m}&
  \colhead{RRL}&
  \colhead{RRL I.D.}&
  \colhead{$<$T$_{\rm C}>$}\\
  \colhead{} &
  \colhead{}&
  \colhead{} &
  \colhead{[$\times10^{-8}$]}&
  \colhead{[$\times10^{-8}$]}&
  \colhead{[$\times10^{-7}$]}&
  \colhead{}&
  \colhead{}&
  \colhead{free--free at 8.5GHz} \\
  \colhead{} &
  \colhead{[\degr]}&
  \colhead{[\degr]} &
  \colhead{[W\,m$^{-2}$\,sr$^{-1}$]}&
  \colhead{[W\,m$^{-2}$\,sr$^{-1}$]}&
  \colhead{[W\,m$^{-2}$\,sr$^{-1}$]}&
  \colhead{[K\,km\,s$^{-1}$]}&
  \colhead{}&
  \colhead{[K]}
  }
\startdata
G302.6+0.0 & 302.553 & 0.0 & 2.352 $\pm$ 0.003 & 2.225 $\pm$ 0.016 & -- & 0.64 $\pm$ 0.09 &H91$\alpha$ & 0.16 $\pm$ 0.00 \\
G305.1+0.0 & 305.106 & 0.0 & 4.656 $\pm$ 0.009 & 6.512 $\pm$ 0.070 & -- & 1.55 $\pm$ 0.07 &H91$\alpha$ & 0.44 $\pm$ 0.00 \\
G306.4+0.0 & 306.383 & 0.0 & 1.012 $\pm$ 0.003 & 1.134 $\pm$ 0.018 & -- & 0.18 $\pm$ 0.03 &H92$\alpha$ & 0.10 $\pm$ 0.00 \\
G307.7+0.0 & 307.660 & 0.0 & 2.467 $\pm$ 0.012 & 2.312 $\pm$ 0.067 & -- & 0.29 $\pm$ 0.01 &H92$\alpha$ & 0.17 $\pm$ 0.00 \\
G310.2+0.0 & 310.213 & 0.0 & 0.999 $\pm$ 0.003 & 0.429 $\pm$ 0.011 & -- & 0.27 $\pm$ 0.02 &H92$\alpha$ & 0.15 $\pm$ 0.00 \\
G314.0+0.0 & 314.043 & 0.0 & 1.885 $\pm$ 0.004 & 1.336 $\pm$ 0.010 & -- & 0.28 $\pm$ 0.03 &H92$\alpha$ & 0.16 $\pm$ 0.00 \\
G316.6+0.0 & 316.596 & 0.0 & 5.633 $\pm$ 0.014 & 4.563 $\pm$ 0.073 & -- & 0.63 $\pm$ 0.03 &H92$\alpha$ & 0.38 $\pm$ 0.00 \\
G317.9+0.0 & 317.872 & 0.0 & 4.064 $\pm$ 0.013 & 3.000 $\pm$ 0.060 & -- & 0.60 $\pm$ 0.03 &H92$\alpha$ & 0.37 $\pm$ 0.00 \\
G326.8+0.0 & 326.808 & 0.0 & 10.015 $\pm$ 0.015 & 6.501 $\pm$ 0.067 & -- & 1.07 $\pm$ 0.07 &H91$\alpha$ & 0.36 $\pm$ 0.00 \\
G330.0+0.0 & 330.000 & 0.0 & 1.470 $\pm$ 0.004 & 0.914 $\pm$ 0.014 & -- & 0.18 $\pm$ 0.09 &H91$\alpha$ & 0.18 $\pm$ 0.01 \\
G331.7+0.0 & 331.739 & 0.0 & 4.953 $\pm$ 0.014 & 3.708 $\pm$ 0.071 & -- & 0.22 $\pm$ 0.06 &H91$\alpha$ & 0.27 $\pm$ 0.02 \\
G332.6+0.0 & 332.609 & 0.0 & 4.583 $\pm$ 0.013 & 2.878 $\pm$ 0.068 & -- & 1.17 $\pm$ 0.17 &H92$\alpha$ & 0.28 $\pm$ 0.02 \\
G333.5+0.0 & 333.478 & 0.0 & 9.275 $\pm$ 0.011 & 5.345 $\pm$ 0.057 & -- & 0.82 $\pm$ 0.12 &H91$\alpha$ & 0.43 $\pm$ 0.02 \\
G336.1+0.0 & 336.087 & 0.0 & 15.073 $\pm$ 0.022 & 10.208 $\pm$ 0.061 & -- & 0.94 $\pm$ 0.16 &H91$\alpha$ & 0.53 $\pm$ 0.03 \\
G337.0+0.0 & 336.957 & 0.0 & 16.668 $\pm$ 0.022 & 11.856 $\pm$ 0.072 & -- & 0.92 $\pm$ 0.13 &H91$\alpha$ & 1.08 $\pm$ 0.07 \\
G337.8+0.0 & 337.826 & 0.0 & 12.177 $\pm$ 0.014 & 8.488 $\pm$ 0.071 & -- & 0.72 $\pm$ 0.11 &H91$\alpha$ & 0.57 $\pm$ 0.04 \\
G338.7+0.0 & 338.696 & 0.0 & 3.371 $\pm$ 0.005 & 2.202 $\pm$ 0.014 & -- & 0.60 $\pm$ 0.09 &H92$\alpha$ & 0.24 $\pm$ 0.02 \\
G342.2+0.0 & 342.174 & 0.0 & 4.645 $\pm$ 0.014 & 4.807 $\pm$ 0.065 & -- & 0.53 $\pm$ 0.08 &H91$\alpha$ & 0.29 $\pm$ 0.02 \\
G343.9+0.0 & 343.913 & 0.0 & 3.718 $\pm$ 0.013 & 2.761 $\pm$ 0.064 & -- & 0.28 $\pm$ 0.08 &H91$\alpha$ & 0.24 $\pm$ 0.02 \\
G345.7+0.0 & 345.652 & 0.0 & 12.527 $\pm$ 0.015 & 5.812 $\pm$ 0.074 & -- & 14.12 $\pm$ 0.28 &H89$\alpha$ & 1.73 $\pm$ 0.07 \\
G346.5+0.0 & 346.522 & 0.0 & 2.879 $\pm$ 0.011 & 1.899 $\pm$ 0.060 & -- & 0.45 $\pm$ 0.08 &H89$\alpha$ & 0.29 $\pm$ 0.02 \\
G349.1+0.0 & 349.130 & 0.0 & 21.023 $\pm$ 0.025 & 9.128 $\pm$ 0.061 & -- & 3.92 $\pm$ 0.05 &H89$\alpha$ & 1.51 $\pm$ 0.06 \\
G350.9+0.0 & 350.870 & 0.0 & 1.410 $\pm$ 0.011 & 1.302 $\pm$ 0.060 & -- & 0.33 $\pm$ 0.09 &H89$\alpha$ & 0.14 $\pm$ 0.01 \\
G353.5+0.0 & 353.478 & 0.0 & 3.672 $\pm$ 0.006 & 2.945 $\pm$ 0.014 & -- & 0.20 $\pm$ 0.07 &H89$\alpha$ & 0.21 $\pm$ 0.02 \\
G354.3+0.0 & 354.348 & 0.0 & 2.396 $\pm$ 0.012 & 1.534 $\pm$ 0.066 & -- & 0.26 $\pm$ 0.04 &H89$\alpha$ & 0.16 $\pm$ 0.01 \\
G355.2+0.0 & 355.217 & 0.0 & 5.180 $\pm$ 0.012 & 1.953 $\pm$ 0.051 & -- & 0.49 $\pm$ 0.04 &H89$\alpha$ & 0.46 $\pm$ 0.06 \\
G356.1+0.0 & 356.087 & 0.0 & 2.596 $\pm$ 0.004 & 1.628 $\pm$ 0.008 & -- & 0.16 $\pm$ 0.04 &H89$\alpha$ & 0.13 $\pm$ 0.02 \\
G359.5+0.0 & 359.500 & 0.0 & 17.209 $\pm$ 0.027 & 9.763 $\pm$ 0.091 & -- & 0.68 $\pm$ 0.15 &H89$\alpha$ & 1.06 $\pm$ 0.01 \\
\enddata
\tablenotetext{}{}
\label{tab:intensities2}
\end{deluxetable*}

The paper is organized as follows. In Section~\ref{sec:observations}
we describe the RRL, [N\,{\sc ii}], and [N\,{\sc iii}] 57$\mu$m
observations.  In Section~\ref{sec:discussion}, we describe our method
to determine the Nitrogen abundance from our observations and we
discuss the properties of the derived Nitrogen abundance
distribution. We also discuss the implications of the observed
distribution in terms of theoretical predictions from chemical
evolution models. In Section~\ref{sec:conclusions} we summarize our
results.

\section{Observations}
\label{sec:observations}

We surveyed the Galactic plane in the Hydrogen radio recombination
line (RRL) emission, using RRL transitions between H89$\alpha$ and
H92$\alpha$, covering 108 lines--of--sights (LOS) in the
$-$135\degr$<l<-$60\degr\ range in Galactic longitude and at a
Galactic latitude of $b=0$\degr, with the NASA DSN DSS--43 70m
telescope and the Green Bank Telescope. These LOS correspond to a
subsample of the {\it Herschel} Open Time Key project, GOT\,C+
\citep{Langer2010,Pineda2013}, which observed the [C\,{\sc ii}]
158$\mu$m line at high spectral resolution with the HIFI
instrument. The GOT\,C+ sample with $b=0$\degr\ was also observed in
[N\,{\sc ii}] 122$\mu$m and 205$\mu$m by \citet{Goldsmith2015}.  The
LOS are sampled every $\sim$1\degr\ in the inner galaxy ($ |l|<60 $
\degr) and every $\sim$2\degr--5\degr\ in the outer galaxy. The
GOT\,C$^+$ goal of adopting this sampling of the Galactic plane was to
obtain a statistical sample of a large number of environments
distributed across the Galaxy.

In the upper panel of Figure~\ref{fig:pv_map} we show the [N\,{\sc
    ii}] 205$\mu$m, 122$\mu$m, and RRL integrated intensities as a
function of Galactic longitude. These intensities are listed in Tables
\ref{tab:intensities} and \ref{tab:intensities2}.  The lower panel of
Figure~\ref{fig:pv_map} shows the longitude--velocity distribution of
the observed RRL emission. The longitude--velocity (LV) map is
overlaid with projections of the Scutum--Crux, Sagittarius--Carina,
Perseus, and Norma-Cygnus Milky Way spiral arms.  For the LV map we
used the fits to the parameters determining the logarithmic spiral
arms presented by \citet{SteimanCameron2010}. We assumed a flat
rotation curve with parameters presented by \citet[][see
  Section~\ref{sec:galact-dist-determ}]{Reid2014}.  We see a good
correspondence between RRL emission and spiral arms, with an enhanced
RRL emission at the spiral arm tangents, which can be explained by the
longer path length for a given velocity range in these regions.

\begin{figure}[t]
\centering
\includegraphics[angle=0,width=0.45\textwidth,angle=0]{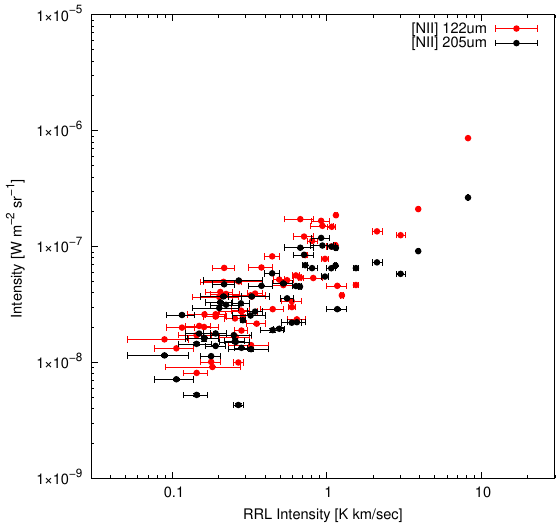}
\caption{[N\,{\sc ii}] 122$\mu$m and 205$\mu$m intensities as a function of the RRL integrated intensity.}
\label{fig:nii_vs_rrl}
\end{figure}

In Figure~\ref{fig:nii_vs_rrl} we show the [N\,{\sc ii}] 122$\mu$m and
205$\mu$m intensities as a function of the RRL integrated intensity
for our sample. The figure shows a linear correlation that suggests
that both [N\,{\sc ii}] and RRL line emission are tracing the same
ionized gas, and that Nitrogen is mostly in a singly ionized form in
our sample, even in the conditions of the Galactic center, whose data
points in the top right of Figure~\ref{fig:nii_vs_rrl}. A more
detailed discussion about the ionization of Nitrogen is presented in
Section~\ref{sec:ioniz-struct-nitr}).

\subsection{Hydrogen Recombination Line Observations}
\label{sec:hydr-recomb-line}

We used the NASA DSN DSS--43 telescope in Camberra, Australia, to
observe 45 LOS in our sample, covering the southern sky portion of the
Galactic plane. We used X--band receiver in position--switching mode
to observe the H91$\alpha$ and H92$\alpha$ hydrogen radio
recombination lines at 8.58482\,GHz and 8.30938\,GHz, respectively.
The angular resolution of the DSS--43 at 8.420\,GHz is 115\arcsec. We
converted the data from an antenna temperature to a main beam
temperature scale using a main beam efficiency of
0.78\footnote{Details in the determination of the aperture efficiency
  and beamsize of the DSS--43 antenna is available on--line at
  \url{https://deepspace.jpl.nasa.gov/dsndocs/810-005/}. }.  Both
lines were resampled to a common spectral grid and averaged together
to increase the signal--to--noise ratio, with the H92$\alpha$
intensities scaled to correspond to that of the H91$\alpha$
lines\footnote{As seen in Equation~(\ref{eq:3}), in LTE, the intensity
  of a RRL is proportional to the $EM$, the line width, and line
  frequency. Given that $EM$ and $\Delta v$ are intrinsic properties
  of the source, the intensity of two RRLs are related by the inverse
  of the ratio of their frequencies \citep{Balser2006}. In the case
  when LTE does not apply (Equation~\ref{eq:7}), however, a dependence
  on electron density and temperature is introduced to the
  relationship between two line intensities. For lines with similar
  principal quantum number, $n$, this effect is negligible, and
  assuming LTE is appropriate. For larger $\Delta n$, however, NLTE
  effects need to be taken into account when scaling RRL intensities.
} .  We fitted a 3rd order polynomial baseline to our data. The
resulting spectra have a typical rms noise of 4\,mK in a
1\,km\,s$^{-1}$ channel. We also re-observed a subsample of LOS in the
H89$\alpha$ (9.17332\,GHz) line with the DSS--43 antenna to improve
the signal--to--noise using the ROACH2 spectrometer
\citep{Virkler2020}. The spectra have a rms noise of 1\,mK over a
3\,km\,s$^{-1}$ channel.

We also observed X--band RRLs in 79 LOS in our sample, covering the
northern sky portion of the Galactic plane, using the Versatile GBT
Astronomical Spectrometer (VEGAS) on the Green Bank Telescope (GBT) in
the position--switching mode.  The angular resolution of the GBT in
X--band is 84\arcsec. For each observed direction, we simultaneously
measured seven H$n\alpha$ RRL transitions in the 9\,GHz band,
H87$\alpha$ to H93$\alpha$, using the techniques discussed in
\citet{Bania2010}, \citet{Anderson2011}, and \citet{Balser2011}, and
averaged all spectra together to increase the signal-to-noise ratio
using {\tt TMBIDL} \citep{Bania2016}. The data were resampled to a
common grid, intensities scaled to correspond to that for the
H89$\alpha$ line (9.17332\,GHz), and we averaged all lines in the band
(2 polarizations per transition) together to increase the
signal--to--noise ratio. The GBT data was calibrated using a noise
diode of known power.  The spectra were later corrected with a 3rd
order polynomial baseline and smoothed to $\sim$1.9\,km\,s$^{-1}$. We
converted the intensities from an antenna temperature to main beam
temperature using a main beam efficiency of 0.94. The typical rms
noise of these data is 2.5\,mK in a 1.9 km\,s$^{-1}$ channel.

\subsection{[N\,{\sc ii}] 122$\mu$m and 205$\mu$m observations }
\label{sec:n-sc-ii}

We used {\it Herschel}/PACS data of the [N\,{\sc ii}] 205$\mu$m and
122$\mu$m lines that were surveyed by \citet{Goldsmith2015} in 149
GOT\,C+ LOS with $b=0$\degr. We refer to \citet{Goldsmith2015} for the
details on the reduction of this data set. The PACS instrument has a
5$\times$5 pixel grid, with a pixel separation of $\sim$9.4\arcsec\,
corresponding to a footprint of 47\arcsec\ in the sky.  The PACS
spectrometer has a resolving power of 1000 at 122$\mu$m and 2000 at
205$\mu$m, corresponding to a velocity resolution of 300\,km\,s$^{-1}$
and 150\,km\,s$^{-1}$, respectively, and therefore the emission lines
are spectrally unresolved. The {\it Herschel} telescope with PACS has
a FWHM beam width of 10\arcsec\ at 122$\mu$m and 15\arcsec\ at
205$\mu$m.

To improve the signal--to--noise of the observations, and to reduce
the contrast between the angular resolution of the [N\,{\sc ii}] and
RRL data sets, we used the footprint--averaged spectra, corresponding
to an angular resolution of 47\arcsec.  (A discussion about beam
dilution effects is presented in
Section~\ref{sec:electr-temp-determ}.)  Note that there is a slight
difference between the [N\,{\sc ii}] 122$\mu$m and 205$\mu$m
intensities we calculated and those presented by
\citet{Goldsmith2015}, as the latter used a simple average of all
pixels in the PACS footprint, while we used a Gaussian--weighted
average based on the distance of each pixel to the center of the
footprint. The difference between these two approaches is minimal and
does not significantly affect the derived electron densities and N$^+$
column densities.  The typical rms noise of the observations are
$\sim2\times10^{-7}$\, erg\,cm$^{-2}$\,s$^{-1}$\,sr$^{-1}$ for the
122$\mu$m line and
$\sim6\times10^{-7}$\,erg\,cm$^{-2}$\,s$^{-1}$\,sr$^{-1}$ for the
205$\mu$m line.

\subsection{[N\,{\sc iii}] 57$\mu$m  observations }
\label{sec:n-sc-iii}

To study the ionization structure of Nitrogen, and to estimate the
contribution of highly ionized states of Nitrogen to the total
Nitrogen abundance, we observed a subsample of 8 LOS in the [N\,{\sc
    iii}] 57$\mu$m line with the Far Infrared Field-Imaging Line
Spectrometer (FIFI--LS; \citealt{Fischer2018}) which is an integral
field far--infrared spectrometer on SOFIA. These observations were
taken as part of the SOFIA ID {\tt 08\_0023} and {\tt 09\_0150}
projects.

The [N\,{\sc iii}] 57$\mu$m spectra were obtained with pointed
observations of the FIFI--LS $5\times5$ pixel footprint in the blue
channel ($\mu$m), with an angular resolution of 6\arcsec\ at
57$\mu$m. To improve the signal--to--noise ratio we averaged the
spectra in each footprint, and therefore the angular resolution of the
average corresponds to the 30\arcsec\ size of the FIFI-LS
footprint. The spectral resolution of FIFI-LS at 57$\mu$m is
280\,km\,s$^{-1}$. Given that the typical line width of the RRL
emission in our sample is 25\,km\,s$^{-1}$, the lines observed with
FIFI--LS are spectrally unresolved.

We computed the footprint--averaged spectrum for each LOS using the
SOSPEX software \citep{Fadda2018} and imported the resulting spectra
into GILDAS/CLASS for baseline corrections and fitting.  The averaged
FIFI--LS spectra at 57\,$\mu$m shows a steep baseline that results
from the effects of rapidly changing atmosphere conditions during
flight. The shape of this baseline is similar to that seen for the
transmission curve of the atmosphere.  We fitted a 3rd order
polynomial to the flux spectrum, using our knowledge of the location
of the spectral lines from the spectrally resolved RRL spectra to
define a window in the flux spectrum where the channels are excluded
from the fit.  The typical rms noise of the observations is
6.51$\times$10$^{-5}$\, erg\,s$^{-1}$\,cm$^{-2}$\,sr. We detected the
[N\,{\sc iii}] 57$\mu$m line in 4 out of the 10 LOS. For the
undetected LOS, we will use their 3$\sigma$ upper limits to constrain
the contribution from doubly ionized Nitrogen in these locations.

\begin{figure}[t]
\centering
\includegraphics[angle=0,width=0.25\textwidth,angle=0]{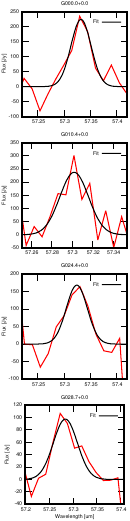}
\caption{Spectra of the detected [N\,{\sc iii}] 57$\mu$m line observed
  with the FIFI--LS instrument on SOFIA. We also show the Gaussian fit
  to the data. }
\label{fig:fifils_spectra}
\end{figure}

We complemented our analysis with {\it Herschel}/PACS observations of
the [N\,{\sc iii}] 57$\mu$m, [N\,{\sc ii}] 122$\mu$m, and [O\,{\sc
    iii}] 88$\mu$m and 52$\mu$m, and SPIRE--FTS [N\,{\sc ii}]
205$\mu$m emission observed in the Sagittarius A region in the
Galactic center. These observations were presented by, and the data
reduction is described in, \citet{Goicoechea2013}.
In our analysis we used the PACS footprint averaged spectrum and
therefore its angular resolution is $\sim$47\arcsec.  The SPIRE--FTS
[N\,{\sc ii}] 205\,$\mu$m spectrum presented by \citet{Goicoechea2013}
was computed in a 30\arcsec\ aperture. As with the FIFI--LS
observations, the spectral lines observed with {\it Herschel}/PACS and
SPIRE are unresolved in velocity.  In Figure~\ref{fig:fifils_spectra}
we show the detected SOFIA FIFI--LS [N\,{\sc iii}] 57$\mu$m
spectra. The intensities of the [N\,{\sc iii}] 57$\mu$m, and 3$\sigma$
upper limits, are listed in in Table~\ref{tab:intensities}. We present
an analysis of this data set in Section~\ref{sec:ioniz-struct-nitr}.

\subsection{Sample Location with Respect to Known H II Regions}\label{sec:sample-location-with}

 To evaluate potential sources of uncertainties in the derived
  Nitrogen abundances, such as the presence of doubly ionized Nitrogen
  and beam dilution effects related to the use of observations from
  different telescopes in compact sources, we need to understand the
  nature of the regions we are sampling.  As discussed in
  Section~\ref{sec:observations}, the sample of LOS used in our study
  is drawn from the {\it Herschel} GOT C+ survey which provided a
  uniform sampling of the Galactic plane and therefore it did not
  intentionally targeted to the center of any specific H\,{\sc ii}
  region. \citet{Goldsmith2015} derived electron densities in this
  sample showing typical values of about 30\,cm$^{-3}$ which are
  higher than what is expected for the Warm Ionized Medium (WIM) but
  lower than that of compact H\,{\sc ii} regions. This result suggests
  that the [N\,{\sc ii}] emission detected in the GOT C+ sample arises
  from an extended component of the ionized ISM, which is not closely
  associated with massive stars. 

To further assess the nature of the sources in our sample we studied
the environment traced by our sample LOS by searching for the nearest
known H\,{\sc ii} regions from the Wide-field Infrared Survey Explorer
(WISE) Catalog of Galactic H\,{\sc ii} Regions
\citep{Anderson2014}. These regions were followed up with RRL and
radio continuum observations to confirm that the mid--infrared warm
dust emission is associated with ionized gas
\citep{Bania2010,Anderson2011,Wenger2019}. In Tables~\ref{tab:xmatch3}
and \ref{tab:xmatch3_v2}, we list the nearest WISE H\,{\sc ii} region
to each LOS in our sample, the distance between the center of the
observations' beam and that of the nearest H\,{\sc ii} region, the
radius of the nearest H\,{\sc ii} region, and the percentage of the
RRL beam area that overlaps with the H\,{\sc i} region.  The radius of
a WISE H\,{\sc ii} region is defined by that of a circular aperture
that encloses its associated mid--infrared emission.  We find that the
majority of LOS in our sample (40 out of 61) do not overlap with kwown
H\,{\sc i} regions. There are, however, 12 LOS in which the RRL beam
has an overlap with the nearest H\,{\sc ii} region of over 50\%, of
which 7 have a 100\% overlap.

 We compared the average electron temperatures, volumen densities, and
 ionized Nitrogen abundances ($12+\log(N^+/H^+)$) for the sample with
 0\% overlap and that with 100\% overlap. These average quantities are
 8500\,K, 25.8\,cm$^{-3}$, and 7.95, for the sample without overlap,
 and 8450\,K, 30.9\,cm$^{-3}$, and 7.85, for the sample that fully
 overlaps with known H\,{\sc ii} regions. We find no significant
 difference in the derived quantities depending on whether they
 overlap with a known H\,{\sc ii} region or not, suggesting that most
 of the sources in our sample are not associated with compact H\,{\sc
   ii} regions and are part of an extended, low ionization gas
 component at the outskirts of H\,{\sc ii} regions.


\begin{deluxetable}{lcrrr} 
\tabletypesize{\footnotesize} \centering \tablecolumns{5} \small
\tablewidth{0pt} \tablecaption{Nearest known H\,{\sc ii} region to sample LOS  for $l\geq0$\degr.} \tablenum{3}
\tablehead{
  \colhead{LOS} &
  \colhead{H\,{\sc ii} Source} &
  \colhead{Distance} &
  \colhead{H\,{\sc ii} Radius} &
  \colhead{Beam overlap} \\
  \colhead{}  & \colhead{} &   \colhead{\arcsec} &    \colhead{\arcsec} & \colhead{\%}}
\startdata
G000.0+0.0 & G000.008+00.036 & 134.9 & 25.0 &    0 \\ 
G000.5+0.0 & G000.522-00.011 & 90.5 & 27.0 &    0 \\ 
G003.5+0.0 & G003.462-00.014 & 75.1 & 16.7 &    0 \\ 
G004.3+0.0 & G004.283+00.031 & 257.1 & 42.5 &    0 \\ 
G005.2+0.0 & G005.186-00.083 & 318.5 & 21.6 &    0 \\ 
G006.1+0.0 & G006.057-00.033 & 158.8 & 42.5 &    0 \\ 
G007.8+0.0 & G007.761+00.006 & 233.7 & 140.1 &    0 \\ 
G008.7+0.0 & G008.685-00.047 & 171.3 & 287.8 &  100 \\ 
G010.4+0.0 & G010.441+00.012 & 51.9 & 53.6 &   44 \\ 
G011.3+0.0 & G011.274-00.053 & 217.9 & 22.5 &    0 \\ 
G012.2+0.0 & G012.145-00.001 & 104.0 & 402.9 &  100 \\ 
G013.0+0.0 & G013.135+00.058 & 391.5 & 55.5 &    0 \\ 
G013.9+0.0 & G013.899-00.014 & 69.3 & 60.0 &   30 \\ 
G014.8+0.0 & G014.867+00.060 & 374.4 & 23.2 &    0 \\ 
G016.5+0.0 & G016.560+00.002 & 141.4 & 35.5 &    0 \\ 
G020.0+0.0 & G019.986+00.094 & 342.6 & 29.3 &    0 \\ 
G020.9+0.0 & G020.959+00.055 & 380.7 & 47.7 &    0 \\ 
G021.7+0.0 & G021.746-00.035 & 125.0 & 85.4 &    1 \\ 
G023.5+0.0 & G023.459-00.026 & 115.5 & 50.5 &    0 \\ 
G024.3+0.0 & G024.356+00.048 & 176.9 & 75.5 &    0 \\ 
G025.2+0.0 & G025.179+00.038 & 194.4 & 137.2 &    0 \\ 
G026.1+0.0 & G026.091-00.057 & 204.1 & 25.6 &    0 \\ 
G027.0+0.0 & G026.984-00.062 & 243.5 & 155.9 &    0 \\ 
G028.7+0.0 & G028.702+00.014 & 57.2 & 70.4 &   63 \\ 
G030.0+0.0 & G030.014+00.017 & 80.5 & 23.7 &    0 \\ 
G031.3+0.0 & G031.264+00.031 & 123.4 & 71.0 &    0 \\ 
G036.4+0.0 & G036.380+00.014 & 53.2 & 53.2 &   42 \\ 
G037.7+0.0 & G037.691+00.027 & 151.1 & 40.4 &    0 \\ 
G041.5+0.0 & G041.512+00.021 & 113.8 & 86.2 &   10 \\ 
G044.0+0.0 & G044.007-00.016 & 138.6 & 27.1 &    0 \\ 
G045.3+0.0 & G045.453+00.044 & 508.0 & 244.8 &    0 \\ 
G049.1+0.0 & G049.163-00.066 & 243.0 & 203.6 &    1 \\ 
G054.3+0.0 & G054.376-00.050 & 469.7 & 121.9 &    0  \\
\enddata
\tablenotetext{}{}
\label{tab:xmatch3}
\end{deluxetable}

\begin{deluxetable}{lcrrr} 
\tabletypesize{\footnotesize} \centering \tablecolumns{4} \small
\tablewidth{0pt} \tablecaption{Nearest known H\,{\sc ii} region to sample LOS for $l<0$\degr} \tablenum{4}
\tablehead{
  \colhead{LOS} &
  \colhead{H\,{\sc ii} Source} &
  \colhead{Distance} &
  \colhead{H\,{\sc ii} Radius} &
  \colhead{Beam overlap}  \\
\colhead{}  & \colhead{} &   \colhead{\arcsec} &    \colhead{\arcsec} & \colhead{\%}}
\startdata
G302.6+0.0 & G302.586-00.029 & 156.6 & 64.2 &    0  \\ 
G305.1+0.0 & G305.201+00.009 & 343.6 & 27.6 &    0  \\ 
G306.4+0.0 & G306.361-00.291 & 1050.1 & 84.1 &    0 \\ 
G307.7+0.0 & G307.850+00.015 & 686.1 & 28.9 &    0  \\ 
G310.2+0.0 & G310.227-00.023 & 96.9 & 325.3 &  100  \\ 
G314.0+0.0 & G314.077+00.004 & 123.5 & 38.2 &    0  \\ 
G316.6+0.0 & G316.548-00.003 & 173.1 & 134.5 &    9 \\ 
G317.9+0.0 & G317.870-00.008 & 26.9 & 132.4 &  100 \\ 
G326.8+0.0 & G326.951+00.009 & 516.1 & 440.1 &    0 \\ 
G330.0+0.0 & G329.976-00.002 & 86.5 & 121.4 &   83 \\ 
G331.7+0.0 & G331.760+00.064 & 242.6 & 109.7 &    0 \\ 
G332.6+0.0 & G332.718-00.053 & 435.7 & 25.2 &    0 \\ 
G333.5+0.0 & G333.580+00.058 & 423.4 & 293.0 &    0 \\ 
G336.1+0.0 & G336.097+00.005 & 40.9 & 219.4 &  100 \\ 
G337.0+0.0 & G336.969-00.013 & 61.7 & 17.3 &    3 \\ 
G337.8+0.0 & G337.827+00.056 & 202.8 & 135.5 &    0 \\ 
G338.7+0.0 & G338.596-00.007 & 360.8 & 31.7 &    0 \\ 
G342.2+0.0 & G342.120+00.001 & 194.5 & 295.2 &  100 \\ 
G343.9+0.0 & G343.912+00.116 & 418.5 & 91.4 &    0 \\ 
G345.7+0.0 & G345.651+00.015 & 55.5 & 98.2 &   90 \\ 
G346.5+0.0 & G346.529-00.013 & 52.2 & 79.2 &   72 \\ 
G349.1+0.0 & G349.126+00.010 & 40.7 & 98.9 &  100 \\ 
G350.9+0.0 & G350.850-00.040 & 159.2 & 43.5 &    0 \\ 
G353.5+0.0 & G353.547-00.013 & 252.4 & 123.4 &    0 \\ 
G354.3+0.0 & G354.356+00.000 & 28.9 & 41.5 &   45 \\ 
G355.2+0.0 & G355.221-00.015 & 53.6 & 99.8 &   93 \\ 
G356.1+0.0 & G356.090-00.075 & 269.6 & 71.9 &    0 \\ 
G359.5+0.0 & G359.486+00.028 & 115.3 & 15.9 &    0 \\
\enddata
\tablenotetext{}{}
\label{tab:xmatch3_v2}
\end{deluxetable}

\begin{deluxetable*}{lcccc} 
\tabletypesize{\footnotesize} \centering \tablecolumns{5} \small
\tablewidth{0pt}
\tablecaption{Summary of Available Radio Continuum Surveys} \tablenum{5}
\tablehead{
\colhead{I.D.} &
\colhead{Frequency} &
\colhead{Angular Resolution} &
\colhead{Galactic Longitude Coverage }&
\colhead{Reference}\\
\colhead{} &
\colhead{[GHz]} &
\colhead{[\arcmin]} &
\colhead{}&
\colhead{}
}
\startdata
VLA Galactic Center & 0.332   &     0.72 &  358\degr$< l < $2\degr\    & \citet{LaRosa2000}\\ 
HASLAM              & 0.408   &       51    & Full Sky                    & \citet{Haslam1981}\\
VGPS                & 1.4     &        1     & 18\degr $< l < $ 67\degr\   & \citet{Stil2006}  \\
SGPS                & 1.4     &      1.6   & 253\degr $< l < $ 358\degr\ & \citet{Haverkorn2006} \\
Effelsberg/GLOSTAR  & 4.88    &      2.4   & 358\degr $< l < $ 60\degr\  & \citet{Brunthaler2021} \\
Parkes 64--m        & 5.0     &      4.1   & 190\degr $< l < $ 40\degr\  & \citet{Haynes1978} \\
Effelsberg/GLOSTAR  & 6.82    &      1.8   & 358\degr $< l < $ 60\degr\  & \citet{Brunthaler2021} \\
Nobeyama 45--m      & 10.3    &      2.66   & 355\degr $< l < $ 56\degr\  & \citet{Handa1987}      \\
\enddata
\tablenotetext{}{}
\label{tab:radio_surveys}
\end{deluxetable*}

\section{Discussion}
\label{sec:discussion}

\subsection{Nitrogen  Abundance Determination}
\label{sec:nitr-abund-determ}

The ionized Nitrogen abundance, $X_{\rm N^+}$, relative to ionized
hydrogen is given by the ratio of the ionized Nitrogen, $N({\rm
  N^+})$, and ionized hydrogen, $N({\rm H^+})$, column densities (see
also \citealt{Pineda2019}),
\begin{equation}\label{eq:1}
    X_{\rm N^+}=\frac{N({\rm N^+})}{N({\rm H^+})}.
\end{equation}
We derived the ionized Nitrogen column density from the [N\,{\sc ii}]
205$\mu$m line intensity ($I_{\rm 205 \mu m})$ using \citep{Goldsmith2015},
\begin{equation}\label{eq:2}
N({\rm N^+})= \frac{4\pi I_{\rm 205 \mu m}}{A_{10} h \nu_{\rm 205 \mu m}
  f_1(n_e,T_e)},
\end{equation}
where the spontaneous decay rate (Einstein's $A$ coefficient) is
$A_{10}=2.08\times10^{-6}$\,s$^{-1}$, the rest frequency is
$\nu_{205\mu m}=1.461\times10^{12}$\,Hz. The fractional population of
the $^3{\rm P}_1$, $f_1$, is a function of the electron density,
$n_e$, and the electron temperature $T_e$.

In local thermodynamical equilibrium (LTE), the main--beam temperature
(in units of K) per unit velocity (km s$^{-1}$) of a hydrogen
recombination line is related to the emission measure, EM, (in units
of cm$^{-6}$\,pc), as \citep{Tools},
\begin{equation}\label{eq:3}
\int T^{\rm RRL}dv=5.76\times10^{11} T_{\rm e}^{-3/2} EM \nu_{\rm RRL}^{-1}, 
\end{equation}
where the speed of light, $c$, is in units of km s$^{-1}$, the rest
frequency of the RRL, $\nu_{\rm RRL}$, in Hz, and the electron
temperature, $T_e$, is in K. The EM is defined as the integral of the
electron volume density squared along the line of sight,
\begin{equation}\label{eq:4}
EM=\int n^2_e dl.
\end{equation}
Assuming that the electron density is constant along the line of
sight, which is approximately valid for the discrete sources we
typically detect, this equation can be simplified to
\begin{equation}\label{eq:5}
EM=n_eN_e\simeq n_eN({\rm H^+}).
\end{equation}
We can thus, re--order equation (\ref{eq:3}) in terms of the H$^+$
column density and electron density as,
\begin{equation}\label{eq:6}
N({\rm H }^+)=\int T^{\rm RRL}dv /(1.87\times10^{-7} \nu_{RRL}^{-1} T_{\rm e}^{-3/2} n_{\rm e}) , 
\end{equation}
where $n_{\rm e}$ is in units of cm$^{-3}$ and $N({\rm H }^+)$ in units of cm$^{-2}$.

The hydrogen recombination line emission can be affected by deviations
from local thermodynamical equilibrium and in this situation this
deviation can be defined in terms of the ratio \citep{Gordon2002},
\begin{equation}\label{eq:7}
G_{\rm LTE}(n_e, T_c)=\frac{ T^{\rm RRL}}{T^{\rm RRL}_{\rm LTE}}=b_n \left [1-\frac{1}{2}\tau_{\rm c} \beta_n  \right ],
\end{equation}
where $b_n$ and $\beta_n$ are the departure coefficient and
amplification factor for a transition with principal quantum number
$n$, respectively, and $T_c$ and $\tau_c$ are the continuum brightness
temperature and opacity, respectively, at $\nu_{RRL}$. The continuum
opacity can be derived from observations of $T_c$, and the electron
temperature, using $\tau_c=T_{\rm c}/T_{\rm e}$.  The effects of
deviations from local thermodynamical equilibrium are well understood
\citep{Gordon2002} and a correction for these effects can be readily
applied.  We evaluated the brightness temperature of the continuum at
the frequency of the RRL observations by extrapolating the synchrotron
and free--free spectral energy distributions from the respective
brightness temperature derived for each LOS, as described in
Section~\ref{sec:electr-temp-determ}.

The electron density can be calculated from the [N\,{\sc ii}] 205
$\mu$m/122 $\mu$m intensity ratio and the electron temperature using
Equations 21 and 22 in \citet{Goldsmith2015}, for a range between 10
and 1000 cm$^{-3}$. The uncertainties in the determination of the
electron density are determined by the uncertainty in the [N\,{\sc
    ii}] 205$\mu$m/122$\mu$m intensity ratio and in on those of the
electron temperature.

\subsubsection{Electron Temperature Determination}
\label{sec:electr-temp-determ}

The electron temperature in optically thin, ionized gas regions that
are in LTE can be derived from the ratio of radio recombination line
(RRL) emission to thermal free-free radio continuum emission. This
derivation is possible because RRL emission is proportional to the
product of the emission measure ($EM$) and temperature to the power of
$-$2.5, while thermal radio free-free emission is proportional to $EM$
times temperature to the power of $-1.35$ \citep{Tools}. As a result,
the ratio of RRL to thermal radio continuum emission is proportional
to temperature to the power of $-1.15$ and is independent of $EM$
(e.g., \citealt{Balser2015}). The electron temperature is therefore
given by,

\begin{equation}
  \begin{aligned}
    \frac{T_e}{K} &=   \Bigl  [ 6.985 \times 10^3 \left(\frac{T_{\rm b}}{T_{\rm L}} \right) \left(\frac{\nu_{rc}}{\rm GHz} \right)^{2.1} \\
      &\cdot   (\nu_{\rm RRL}^{-1}) \cdot (\Delta v)^{-1} \cdot (1+y)^{-1}  \Bigr
    ]   ^{0.86956},
  \end{aligned}\label{eq:8}
\end{equation}

where $T_b$ and $T_L$ are the brightness temperatures of the free-free
continuum and RRL peak intensity, respectively, $\Delta v$ is the RRL
full width at half maximum, $\nu_{rc}$ is the frequency of the radio
continuum emission, $\nu_{RRL}$ is the frequency of the RRL
observations, and $y$ is a term related to the contribution of
$^4$He$^+$, which is assumed to be 0.08 \citep{Balser2011}.

There are several radio continuum surveys with a spatial coverage that
overlaps that from our RRL survey and that have similar angular
resolution, so that we can extract intensities for our analysis. These
surveys include the VLA Galactic Plane Survey \citep{Stil2006}, the
Southern Galactic Plane Survey \citep{Haverkorn2006}, the Nobeyama
Radio Observatory 45--m telescope survey \citep{Handa1987}, the Parkes
64--m telescope 6\,cm survey\footnote{Data from the Nobeyama 45--m and
  Parkes 64--m surveys are, among other Galactic plane surveys,
  available for download at the MPIFR's survey sampler at
  \url{https://www3.mpifr-bonn.mpg.de/survey.html}. }
\citep{Haynes1978}, and the Effelsberg 100--m part of the GLOSTAR
survey \citep{Medina2019,Brunthaler2021}.  In
Table\,\ref{tab:radio_surveys} we list the frequency, angular
resolution and Galactic longitude coverage of these surveys.  These
surveys have varying angular resolutions and frequencies, and thus
uncertainties in the relative calibration and the correction from the
contribution from synchrotron emission can vary from survey to survey.
To minimize these uncertainties, in each LOS we corrected all
available continuum brightness temperatures for the contribution from
synchrotron emission at their frequencies (see below), and estimated
the continuum brightness temperatures at the frequency of our RRL
observations, 8.5\,GHz, assuming a free-free spectrum with spectral
index of $ -2.1$. We then averaged all available samples together to
obtain an average free--free brightness temperature at 8.5\,GHz, which
are listed in Tables\,\ref{tab:intensities} and
\,\ref{tab:intensities2}. In our analysis, we only used radio
continuum brightness temperatures with a signal--to--noise ratio
larger than 10.  In Appendix~\ref{sec:depend-nitr-abund} we show the
N$^{+}$/H$^{+}$ distribution as a function of Galactocentric distance
derived using electron temperatures derived from each of the radio
continuum surveys listed in Table\,\ref{tab:radio_surveys}
individually, showing that its distribution is not significantly
affected by the choice of radio continuum survey used to derive
electron temperatures.

We estimated the contribution of synchrotron emission at a given
frequency using the 408\,MHz map from \citet{Haslam1981} and assuming
a synchrotron spectral index, in brightness temperature scale, of $
-2.8$. The 408\,MHz map has an angular resolution of 51\arcmin\ which
is significantly larger than that of our observations. Note however,
that synchrotron emission in the Galactic plane is expected arise from
diffuse spatially extended gas, while free--free emission originates
from more compact and denser regions, so that we expect that
uncertainties related to the difference in angular resolution are not
significant. In the LOS toward the Galactic center region, where
several compact non-thermal features are observed, we used the VLA
332\,MHz map presented by \citet{LaRosa2000} convolved the angular
resolution of our observations.  We find that the typical contribution
from synchrotron to the observed radio continuum brightness
temperature in our sample is $\sim$47\% at 1.4\,GHz, $\sim$40\% at
5\,GHz, $\sim$36\% at at 6.82\,GHz, $\sim$26\% at 10.3\,GHz.

As mentioned above, sources of uncertainties in using this derivation
of the electron temperature include measurement uncertainties of the
continuum and RRL emission, calibration uncertainties between the
different frequency bands, the relative contribution from synchrotron
and free-free emission to the observed continuum emission, and non-LTE
effects for the RRL intensities.  Note that pressure broadening is not
expected to be significant in the density regime that we are sampling
(\citealt{Brocklehurst1972}).  We used the electron densities from the
[N\,{\sc ii}] lines to account for non--LTE effects on $T_L$ using
equation~(\ref{eq:7}).  In Appendix B, we compare our methodology to
derive electron temperature against electron temperatures derived in a
sample of H\,{\sc ii} regions by \citet{Balser2015} in which
calibration uncertainties and synchrotron contribution are carefully
assessed. We find that for electron temperatures derived using
continuum at both 1.4\,GHz and 6.82\,GHz there is a scatter of about
20--30\% which we attribute to the unaccounted uncertainties described
above.

We also studied whether beam dilution effects resulting from using
observations with different angular resolution in the RRL and
continuum emission impact our determination of electron temperatures
in our sample.To study beam dilution effects in the northern part of our sample, we
smoothed the 60\arcsec\ VGPS survey at 1.4\,GHz to the 84\arcsec\ and
160\arcsec\ resolutions of the GBT RRL and Nobeyama Radio Continnum
data sets, respectively, and studied the intensity ratio at these two
different angular resolutions. We found that beam filling effects are
small for our sample, with typical variations smaller than 5\%.
To study beam dilution effects in the southern part of our sample, we
also convolved the SGPS data at 100\arcsec\ to the 115\arcsec\ angular
resolution of the DSS--43 data. We found small variations in the
intensity ratio of the SGPS continuum data at 100\arcsec\ and
115\arcsec\, with typical variations smaller than 2\%. Because the
variation in intensities due to beam filling in our sample are small,
suggesting that most sources are extended, we did not apply a beam
filling correction to our data.

\begin{figure}[h]
\centering
\includegraphics[angle=0,width=0.45\textwidth,angle=0]{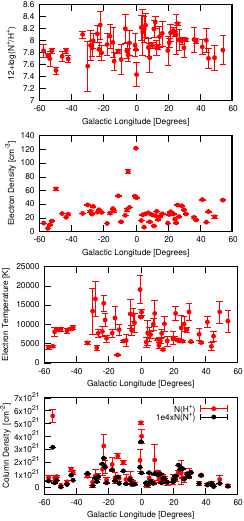}
\caption{Derived ionized Nitrogen abundances, electron densities,
  electron temperatures, and N$^+$ and H$^+$ column densities as a
  function of Galactic longitude.  }\label{fig:results}
\end{figure}

 In Tables~\ref{tab:results1} and \ref{tab:results2} we show the
 derived ionized Nitrogen abundances, electron densities, electron
 temperatures, and N$^+$ and H$^+$ column densities for our sample. In
 Figure~\ref{fig:results} we show the distribution of these quantities
 as a function of Galactic longitude. As seen in
 Figure~\ref{fig:results}, the derived electron temperatures range
 between 3500\,K and 21000\,K, with an average value of $T_{\rm e} =
 8225$\,K, which is typical of ionized gas regions, with a standard
 deviation of 3900K.  We notice that there is a dependence in the
 value of $T_{\rm e}$ with the signal--to--noise ratio of the RRL
 observations, with a tendency for temperatures to be higher at lower
 SNR values. The average temperature for SNR$>10$ is 7484\,K, while
 for $5<$SNR$<10$, it is 9343\,K.  This difference in the electron
 temperature has a small impact on the derived ionized Nitrogen
 abundances. Using Equations~(\ref{eq:1}), (\ref{eq:2}), (\ref{eq:6}),
 and (\ref{eq:8}), for the observed range in ionized Nitrogen and RRL
 intensities, we can derive that N$^+$/H$^+$ is proportional to the
 electron temperature between $T^{-1.1}_{\rm e}$ and $T^{-1.2}_{\rm
   e}$.  Because N$^+$/H$^+$ has an additional dependence on the RRL
 line intensity as $T^{-1}_{\rm L}$ (Equation~\ref{eq:6}), and the
 electron temperature depends on the RRL intensity as $T^{-0.87}_{\rm
   L}$ (Equation~\ref{eq:8}), the resulting dependence of N$^+$/H$^+$
 on the RRL intensity is weak ($\sim T^{-0.04}_{L}$). Thus, an
 uncertainty of a factor of 2 arising from $T_{\rm L}$ would impact
 $T_{\rm e}$ by the same factor but N$^+$/H$^+$ by only a factor of
 1.03.  Note however, that N$^+$/H$^+$ is proportional to $\sim T_{\rm
   b}^{-1}$, and therefore uncertainties from the continuum intensity
 can have a larger impact in the derived Nitrogen abundance. This
 result motivated us to adopt a larger SNR$>10$ criterion for
 selecting lines-of-sights from the radio continuum data set.

\subsubsection{The ionization structure of  Nitrogen}
\label{sec:ioniz-struct-nitr}

The ionization potential of Nitrogen (14.53\,eV) is greater than that
of hydrogen (13.60\,eV) by 0.93\,eV, so Nitrogen can be in neutral
form in regions where Hydrogen is fully ionized by photons between
13.60\,eV. and 14.53\,eV. In the ISM, four primary processes
contribute to Nitrogen ionization: EUV (14.53\,eV to 124.24\,eV)
photoionization, electron collisional ionization, proton (H$^+$)
charge transfer, and X-ray photoionization. While electron
recombination with N$^+$ serves as the primary loss mechanism
\citep{Langer2015,Langer2021}. Models show that electron collisional
ionization of Nitrogen alone is inefficient at temperatures below
$\sim$10$^{4}$\,K due to the large ionization potential of atomic
Nitrogen. However, proton charge transfer (H$^+ + $N$ \rightarrow$ H$
+$ N$^{+} -$ 0.93\,eV), which has a smaller energy barrier, might be
important at temperatures above 5000\,K (see Section 4 in
\citealt{Langer2015} and \citealt{Langer2021}). Therefore a
significant fraction of Nitrogen might be ionized even where few EUV
photons are present above 14.54\,eV due to collisional and exchange
ionization by electrons and protons, respectively.  Note, however,
that a medium where Nitrogen is fully ionized is difficult to attain
without EUV photons because collisional ionization with electrons or
charge exchange with protons is generally balanced by electron
recombination and therefore roughly independent of electron density
\citep{Langer2021}. It is only where EUV dominates ionization
that an increase in photon flux can eventually overcome electron
recombination leading to a fully ionized Nitrogen gas.

Under typical ISM and H\,{\sc ii} region conditions, higher
ionization states of Nitrogen, such as N$^{++}$, can be maintained
only by the presence of an EUV field.  Models show that an electron
temperature greater than about 25000\,K is required for collisional
ionization to be important. Thus, under typical temperatures and
densities of the ionized gas, as determined here, N$^{++}$ will not
have a significant abundance without a source of EUV photons, such as
are typically present in the close vicinity of massive stars.  This
result suggests that regions with higher ionization levels than
N$^{+}$, such as can be probed with the [N\,{\sc iii}] 57$\mu$m line,
are likely to be compact and closely associated with H\,{\sc ii}
regions. It is possible, however, that EUV photons leakage from
H\,{\sc ii} regions \citep{Luisi2019} might make a more diffuse,
extended component of highly ionized Nitrogen possible
\citep{Mizutani2002}.

If a significant fraction of the gas--phase Nitrogen is at ionization
levels higher than N$^{+}$, our assumption that the N/H abundance
ratio can be traced by the N$^{+}$/H$^+$ abundance ratio might not be
valid, and therefore it introduces uncertainties into our
analysis. Note that the linear correlation between the [N\,{\sc ii}]
and RRL lines shown in Figure~\ref{fig:nii_vs_rrl} suggest that if
there is any underestimation of the total Nitrogen along the
line--of--sight, the effect is not significant.

To determine whether higher ionization states, such as N$^{++}$, might
be a significant source of Nitrogen in our sample, we used SOFIA and
{\it Herschel} observations of the fine structure line of doubly
ionized Nitrogen [N\,{\sc iii}] 57$\mu$m, to characterize the
ionization environment of 8 LOS in our survey, and in the Sgr A region
in the Galactic center. Under the typical physical conditions of our
LOS sample (temperatures less than 20,000\,K), N$^{++}$ can only be
maintained by EUV photons from massive stars.  Therefore, the presence
of [N\,{\sc iii}] will enable us to determine whether EUV photons play
an active role in determining the ionization structure of Nitrogen in
our sample.

 The [N\,{\sc iii}] 57$\mu$m line was detected in 4 out of the 8 LOS,
 and their intensities, and 3$\sigma$ upper limits in case of the
 non--detections, are listed in Table~\ref{tab:intensities} and the
 detected spectra are shown in Figure~\ref{fig:fifils_spectra}.  As we
 can see in Table~\ref{tab:xmatch3}, these LOS are either inside
 (G010.4+0.0 and G028.7+0.0), or in the close vicinity of (G000.0+0.0
 and G024.3+0.0), dense H\,{\sc ii} regions cataloged with WISE. This
 association suggests that their environments can be influenced by EUV
 photons that can further ionize N$^{+}$ to N$^{++}$.

\begin{figure}[t]
\centering
\includegraphics[angle=0,width=0.45\textwidth,angle=0]{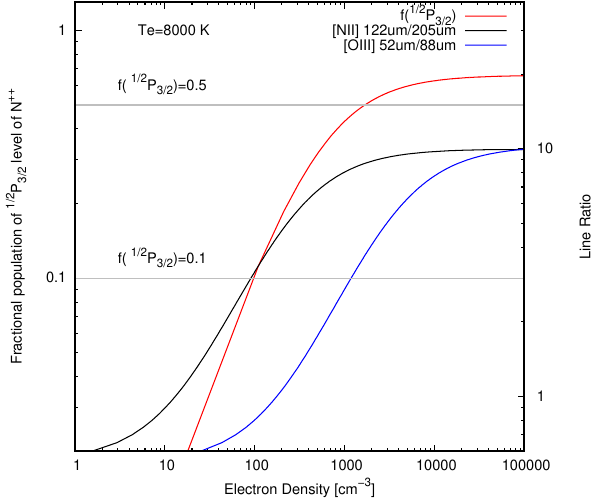}
\caption{Fractional population of the $^{1/2}{\rm P}_{3/2}$ level of
  N$^{++}$, and [N\,{\sc ii}] 205$\mu$m/122$\mu$m and [O\,{\sc iii}]
  52$\mu$m/88$\mu$m line ratios, as a function of the electron density
  for an electron temperature of 8000\,K. The horizontal lines
  highlight the electron densities at which the fractional level
  populations are 10\% and 50\%. }
\label{fig:level_population}
\end{figure}

 In Table~\ref{tab:highly_ionized_gas}, we present the results of our
 analysis of the [N\,{\sc iii}] 57$\mu$m observations. We derived
 N$^{++}$ column densities from the observed [N\,{\sc iii}] 57$\mu$m
 intensity and the electron density derived from the [N\,{\sc ii}]
 122$\mu$m/205$\mu$m ratio using,
\begin{equation}\label{eq:9}
N({\rm N^{++}})= \frac{4\pi I_{\rm 57 \mu m}}{A_{ul} h \nu_{\rm 57
    \mu m} f_{3/2}(n_e)},
\end{equation}
where the spontaneous decay rate for N$^{++}$ is
$A_{ul}=4.79\times10^{-5}$\,s$^{-1}$, and the rest frequency is
$\nu_{57\mu m}=5.229\times10^{12}$\,Hz. The fractional population of
the $^{1/2}{\rm P}_{1/2}$, $f_{3/2}$, is a function of the electron
density, $n_e$, as shown in Figure~\ref{fig:level_population}. The
derived N$^{++}$/N$^+$ ratio for the detected sources ranges from 0.09
to 0.87, suggesting that doubly ionized Nitrogen in these regions is
not dominant, but could be a significant fraction of the total
Nitrogen, and thus can introduce an underestimation of the total
Nitrogen abundance derived from N$^+$/H$^+$ between factors of
$\sim1.09$ for G000.0+0.0 and $\sim1.87$ in G010.4+0.0.  A similar
range is obtained for the 3$\sigma$ upper limits.
However, these results are based on the assumption that the electron
density of the [N\,{\sc iii}]--emitting region is the same as that of
the [N\,{\sc ii}]--emitting region, whereas, given the EUV
requirements to produce N$^{++}$, it is more likely that [N\,{\sc
    iii}] comes from compact regions closer to the source of H\,{\sc
  ii} regions. Thus, under these assumptions, the derived
N$^{++}$/N$^+$ ratio should be considered as an upper limit.  As we
can see in Figure~\ref{fig:level_population}, for the typical
densities derived from the Nitrogen lines ($\sim$20-100\,cm$^{-3}$),
the population of the $^{1/2}{\rm P}_{3/2}$ level is very small
($<10$\%), and thus a relatively large N$^{++}$ column density is
needed to reproduce the observed [N\,{\sc iii}] 57$\mu$m intensities
and upper limits. For $T_{\rm e}$=8000\,K, the $^{1/2}{\rm P}_{3/2}$
level is 50\% populated at about 1000\,cm$^{-3}$, and assuming such a
density in our analysis would result in a N$^{++}$ column density and
N$^{++}$/N$^{+}$ ratio that are a factor of $\sim$10 lower that those
resulting from electron densities determined from the [N\,{\sc ii}]
122$\mu$m/205$\mu$m ratio.  A more appropriate tracer of the volume
density of gas associated with doubly ionized Nitrogen is the [O\,{\sc
    iii}] 52$\mu$m/88$\mu$m ratio, as the critical density of the
[O\,{\sc iii}] 52$\mu$m line is similar to that for [N\,{\sc iii}]
57$\mu$m, as is the EUV requirement to produce O$^{++}$.

The Sagittarius A region in the Galactic center was observed in the
[N\,{\sc ii}] 122$\mu$m and 205$\mu$m and the [O\,{\sc iii}] 52$\mu$m
and 88$\mu$m lines by {\it Herschel}/PACS and represents an ideal
location to study the ionization structure of Nitrogen without
uncertainties in the volume density determination. We used both the
[N\,{\sc ii}] 122$\mu$m/205$\mu$m ratio and the [O\,{\sc iii}]
52$\mu$m/88$\mu$m line ratios to derive the electron volume density in
the low and high ionization regions, respectively. We followed the
procedure discussed in Section~\ref{sec:nitr-abund-determ} and used
the ratio of the X--band continuum brightness temperature determined
from the GBT, and the H92$\alpha$ Hydrogen recombination line data
presented by \citet{Wong2016}, including a correction for the
contribution of Synchrotron emission, to derive an electron
temperature of 11443\,K (see Table~\ref{tab:highly_ionized_gas}).  We
derive an electron density of 254\,cm$^{-3}$ from the [N\,{\sc ii}]
lines, and 2874\,cm$^{-3}$, from the [O\,{\sc iii}] lines. Using the
derived volume densities for low ([N\,{\sc ii}]) and high ([O\,{\sc
    iii}]) ionization regions, and the electron temperature, we
calculated a column density of singly ionized Nitrogen of
$6.3\times10^{17}$\,cm$^{-2}$ and of doubly ionized Nitrogen
$3.7\times10^{16}$\,cm$^{-2}$.  We find that N$^{++}$ represents only
a small fraction (6\%) of the total N$^{+}+$N$^{++}$.  Thus, when
appropriate electron densities are used for the different ionization
regimes, we find that most of the ionized gas mass in this region is
at a low ionization state where most of the Nitrogen is singly
ionized.  Using the parameters derived above, and assuming that the
emission measure from the RRL lines can be separated between that
arising from low and high ionization regions using the 6\% ratio
derived for Nitrogen, we derive an  Nitrogen abundance for this
region of 12+log(N/H)=7.86, which is consistent to those
derived in our sample for the Galactic center.

\begin{deluxetable*}{lccccccc} 
\tabletypesize{\footnotesize} \centering \tablecolumns{4} \small
\tablewidth{0pt}
\tablecaption{Single and Doubly ionized Nitrogen Column Densities} \tablenum{6}
\tablehead{
\colhead{LOS} &
\colhead{$T_{e}$} &
\colhead{$n_{e}$  ([N\,{\sc ii}])} &
\colhead{$N(\rm N^+)$}&
\colhead{$n_{e}$  ([O\,{\sc iii}])} &
\colhead{$N(\rm N^{++})$}&
\colhead{$N(\rm N^{++})/N(\rm N^{+})$}\\
\colhead{} &
\colhead{[K]} &
\colhead{[cm$^{-3}$]} &
\colhead{[10$^{17}$\,cm$^{-2}$]}&
\colhead{[cm$^{-3}$]}&
\colhead{[10$^{17}$\,cm$^{-2}$]}&
\colhead{}&
}
\startdata
 G000.0+0.0  & 11910 $\pm$ 97 & 121.47 $\pm$ 0.52 & 3.58 $\pm$ 0.01 &  -- &  0.32 $\pm$ 0.05 &  0.09 $\pm$ 0.01 \\
 G007.0+0.0  & 8000 $\pm$ 800 & 32.77 $\pm$ 0.66 & 0.81 $\pm$ 0.01 &  -- &  $<$0.43 $\pm$ 0.14 &  $<$0.53 $\pm$ 0.18  \\ 
 G010.4+0.0  & 6129 $\pm$ 367 & 24.32 $\pm$ 0.51 & 1.01 $\pm$ 0.01 &  -- &  0.88 $\pm$ 0.10 &  0.87 $\pm$ 0.10  \\
 G012.2+0.0  & 5800 $\pm$ 648 & 23.35 $\pm$ 0.43 & 1.09 $\pm$ 0.01 &  -- &  $<$0.65 $\pm$ 0.22 &  $<$0.60 $\pm$ 0.20  \\ 
 G023.5+0.0  & 5949 $\pm$ 339 & 25.99 $\pm$ 0.25 & 1.77 $\pm$ 0.01 &  -- &  $<$0.55 $\pm$ 0.18 &  $<$0.31 $\pm$ 0.10  \\ 
 G024.3+0.0  & 5744 $\pm$ 276 & 18.15 $\pm$ 0.29 & 1.45 $\pm$ 0.01 &  -- &  1.15 $\pm$ 0.16 &  0.80 $\pm$ 0.11  \\ 
 G026.1+0.0  & 5630 $\pm$ 198 & 38.83 $\pm$ 0.42 & 1.49 $\pm$ 0.01 &  -- &  $<$0.71 $\pm$ 0.24 &  $<$0.47 $\pm$ 0.16  \\ 
 G028.7+0.0  & 8663 $\pm$ 612 & 37.62 $\pm$ 0.56 & 1.06 $\pm$ 0.01 &  -- &  0.48 $\pm$ 0.05 &  0.45 $\pm$ 0.05  \\
 SGRA  & 9601 $\pm$ 367 & 237.63 $\pm$ 27.43 & 6.34 $\pm$ 0.63 &  2686.6$\pm$1027.9 &  0.37 $\pm$ 0.01 &  0.06 $\pm$ 0.01  \\
\enddata
\tablenotetext{}{}
\label{tab:highly_ionized_gas}
\end{deluxetable*}

\begin{figure}[h]
\centering
\includegraphics[angle=0,width=0.45\textwidth,angle=0]{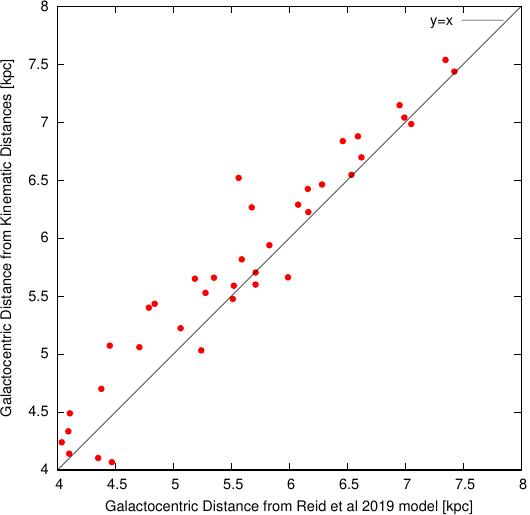}
\caption{Comparison between Galactocentric distances determined using
  the \citet{Reid2019} model and those derived from kinematic
  distances. Only data points with $R_{\rm gal}>4$\,kpc are shown, as
  kinematic distances are not accurate for non-circular motions
  observed in sources associated with the Galactic bar in the
  innermost part of the Galaxy. }\label{fig:bessel_vs_kinematic}
\end{figure}

\begin{figure}[t]
\centering
\includegraphics[angle=0,width=0.47\textwidth,angle=0]{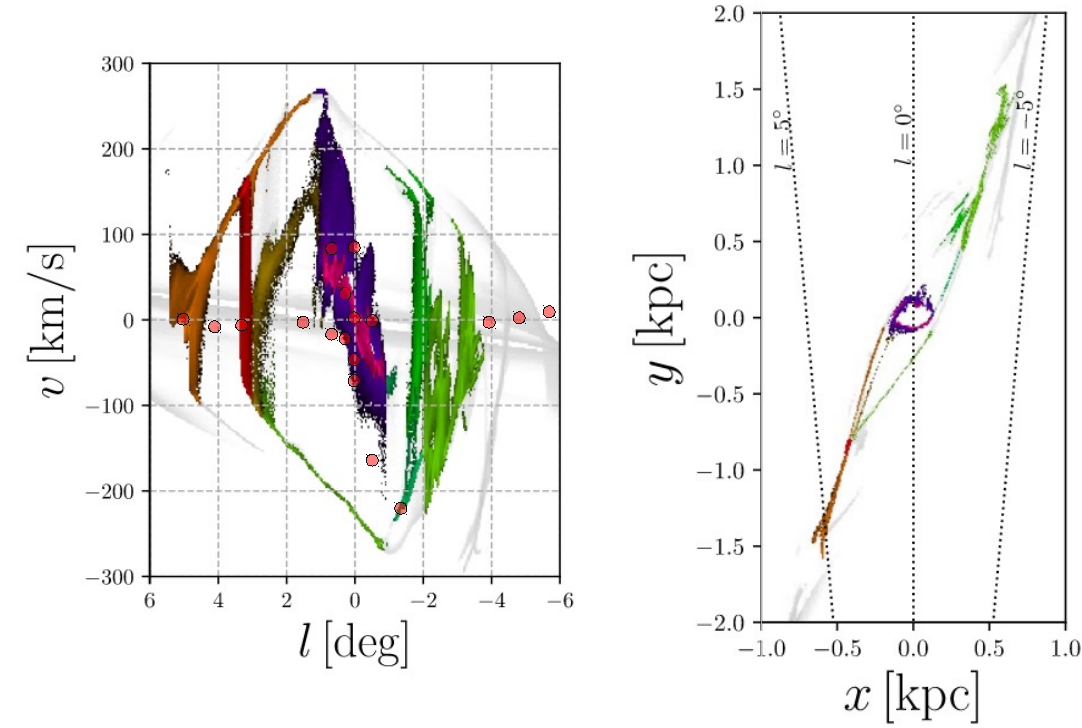}
\caption{Gas kinematics in the longitude--velocity map ({\it left})
  and corresponding spatial distribution ({\it right}) of the Galactic
  center region predicted by numerical simulations of gas flows in a
  barred potential presented by \citet{Sormani2019}.  We overplotted
  the fitted velocity of the RRLs for the 18 LOSs in our survey in
  this region as orange dots.}\label{fig:Galactic_center}
\end{figure}

\begin{figure}[h]
\centering
\includegraphics[angle=0,width=0.5\textwidth,angle=0]{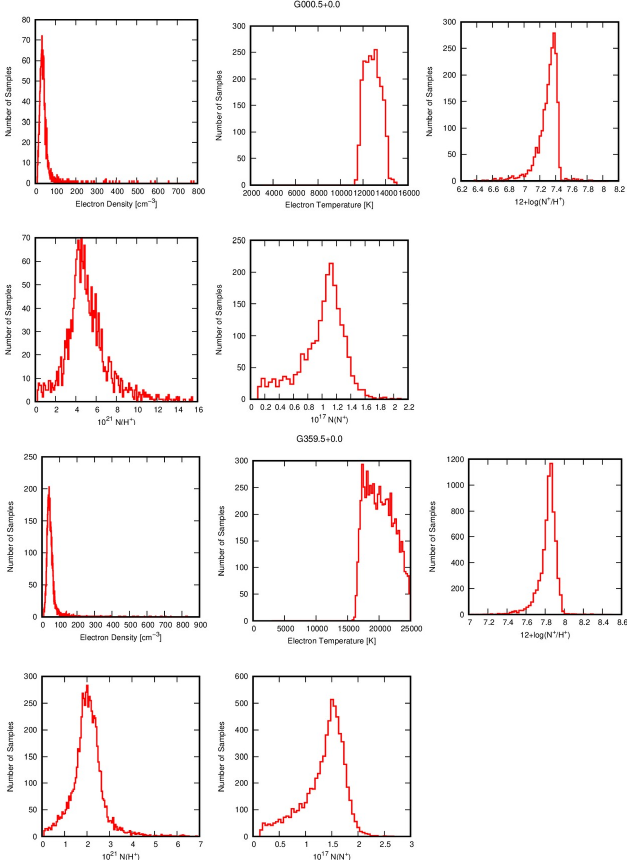}
  \caption{Sample solutions for N$^+$/H$^+$ and constrained values of
    the electron density, electron temperature, and N$^+$ and H$^+$
    column density for G000.5+0.0 and G359.5+0.0 using the procedure
    described in Section~\ref{sec:galact-dist-determ}. }\label{fig:solution_sample}
\end{figure}

\subsubsection{Galactocentric Distance Determination}
\label{sec:galact-dist-determ}

Because our sample in the inner Galaxy is uniformly distributed in
Galactic longitude, we expect that the LOS analyzed here have a wide
range of galactocentric distances in the inner Galaxy, so that we can
study the radial distribution of the ionized Nitrogen abundance in the
Milky Way. The traditional method for determining the Galactocentric
distances from sources in the Galactic plane is the use of kinematic
distances derived from the LSR velocity and Galactic coordinates of
the source.  However, because of non circular motions associated with
the Galactic bar, kinematic distances cannot be accurately determined
in the inner $R_{\rm gal}\lesssim4$\,kpc of the Galaxy.

To ensure that we are able to sample the Galactic plane in the
innermost parts of the Galaxy and the Galactic center, we instead use
the model of the Galaxy presented by \citet[][see also \citealt{Reid2016}]{Reid2019} to determine the
Galactocentric distances to the sources detected in our
survey. \citet{Reid2019} presented a model of the spiral structure of
the Milky Way based on 200 trigonometric parallaxes of masers
associated with massive star forming regions. Distances measured from
maser trigonometric parallaxes have the advantage that they do not
rely on any assumption on the kinematics of the Galaxy. The
\citet{Reid2019} model enables us to estimate distances to sources
using their Galactic coordinates and LSR velocity.  From a distance,
$D$, to a source with Galactic longitude $l$ and latitude $b=0$\degr,
the corresponding Galactocentric distance is given by,
\begin{equation}\label{eq:13}
R_{\rm gal}=\sqrt{D^2-2R_{\odot}D\cos(l)+R_{\odot}^2},
\end{equation}
where $R_{\odot}$ is the distance from the Sun to the Galactic center
which is fitted to be $R_{\odot}$ = 8.15\,kpc by \citet{Reid2019}.
 
In Figure~\ref{fig:bessel_vs_kinematic}, we show a comparison between
Galactocentric distances derived for our sample using the
\citet{Reid2019} model and those derived from kinematic distances, for
$R_{\rm gal}>$4\,kpc. The kinematic distances are determined for a
given velocity component with Galactic longitude $l$, latitude $b$,
and local standard of rest (LSR) velocity $V_{\rm LSR}$, is given by
\begin{equation}\label{eq:10}
R_{\rm gal}= R_{\odot} \sin (l)\cos(b) \left (\frac{V(R_{\rm gal})}{V_{\rm LSR}+V_{\odot}\sin(l)\cos(b)} \right ),
\end{equation}
where $V_{\odot}$ is the orbital velocity of the Sun with respect to
the Galactic center, and $V(R_{\rm gal})$ is the rotation curve.  We
assume a "Universal" rotation curve presented by \citet{Persic1996}
(see Equation 3 in \citealt{Wenger2018}) assuming the value of the
Sun's rotation velocity, $V_{\odot}$ = 247\,km\,s$^{-1}$, fitted by
\citet{Reid2019}.  We tested the dependence of $R_{\rm gal}$ with other
determinations of $V_{\odot}$ \citep{Zhou2023,Akhmetov2024}, finding
negligible differences.  As we can see, the \citet{Reid2019} model is
in very good agreement with Galactocentric distances derived from
kinematic distances, with values of $R_{\rm gal}$ derived from the
\citet{Reid2019} model being on average $\sim$5\% lower than those
derived from kinematic distances.

There are 24 LOS in our sample that have a single RRL velocity
component enabling us to directly associate this emission with the
velocity--unresolved [N\,{\sc ii}] 205$\mu$m and 122$\mu$m and radio
continuum emission. In case of multiple component LOS, we determined
$R_{\rm gal}$ for each velocity component and determined the range of
galactocentric distances that these velocities represent. We assume
that the N$^{+}$/H$^{+}$ gradient is smooth at $<$1.5\,kpc scales and
thus we can assume that the velocity components in given LOS have the
same Nitrogen abundance in case they are at least within 1.5\,kpc from
each other.  There are 14 LOS in this category for which the relative
distance between the two components is lower than 1.5\,kpc.  For this
sub sample, we assigned the average radial distance to the derived
ionized Nitrogen abundance and considered the radial range as error
bars in the X--axis.

There is a subset of 8 LOS, mostly located in the inner
$\pm$20\degr\ from the Galactic center, where a high LSR velocity
component with $v_{\rm LSR}>\pm$80\,km\,s$^{-1}$ together with another
low LSR velocity at about $v_{\rm LSR}\simeq $0\,km\,s$^{-1}$, are
observed.  The \citet{Reid2019} model suggests that the low velocity
components are at distances larger than 4\,kpc from the Galactic
Center while the components with high LSR velocity are in the
0\,kpc$<R_{\rm gal}<4$\,kpc range. Thus, the high LSR velocity
components are likely associated with the Galactic bar and located in
the proximity of the Galactic center.
In Figure~\ref{fig:Galactic_center}, we show the result of numerical
simulations of gas flows in a barred potential presented by
\citet{Sormani2019}, with the left panel showing the predicted gas
kinematics in the longitude--velocity map of this region and in the
right panel the spatial distribution of these different components. We
overplotted the fitted velocity of the RRLs for the 18 velocity
components in our survey in this region as orange dots. As we can see
the high LSR velocity components are likely associated with the bar
potential gas corresponding to the inner 2\,kpc of the Galaxy. At the
same time, sources with velocities near $v_{\rm LSR} \simeq$
0\,km\,s$^{-1}$ are likely at higher Galactocentric distances. In the
figure, we see that the G000.0+0.0 LOS shows multiple velocity
components which are predicted to be associated with the Galactic
center. We therefore assume for this LOS that the N$^{+}$/H$^{+}$,
electron temperature, and electron density of the gas is the same for
all velocity components, and therefore we can combine the RRL data
with the unresolved PACS and continuum data to derive N$^{+}$/H$^{+}$.

Because the observed velocity unresolved [N\,{\sc ii}] 122$\mu$m and
205$\mu$m emission is the sum of the intensity of these lines arising
from both velocity components, to determine the N$^{+}$/H$^{+}$ in the high LSR
velocity sources we need to estimate the contribution to the observed
[N\,{\sc ii}] intensity from the low LSR velocity sources.  As
discussed above, the \citet{Reid2019} model suggests that the low
velocity components are at distances larger than 4 kpc from the
Galactic Center. If that is the case, we can assume that the N$^{+}$/H$^{+}$ for
the low LSR velocity sources is within the range observed for all
other sources across the Galaxy with $R_{\rm gal}>4$\,kpc, and we can
determine its value using the fit to the N$^{+}$/H$^{+}$ distribution as a
function of Galactocentric distance derived below
(Equation~\ref{eq:12}).  With an assumed N$^{+}$/H$^{+}$, the intensity of the
[N\,{\sc ii}] 122$\mu$m and 205$\mu$m emission for the low velocity
component is given by,
\begin{equation}\label{eq:11}
  I^{low}_{122\mu m}=  \frac{A_{21}h\nu_{122\mu m}X_{\rm N^+}EMf_2(ne,T_e)}{4\pi ne}
\end{equation}
and
\begin{equation}\label{eq:14}
 I^{low}_{205\mu m}=  \frac{A_{10}h\nu_{205\mu m}X_{\rm N^+}EMf_1(ne,T_e)}{4\pi ne}
\end{equation}
where $f_1$ and $f_2$ are the level populations of the $^3{\rm P}_1$
and $^3{\rm P}_2$ levels, respectively, and the EM is derived from the
RRL observations for this velocity component using
Equation~(\ref{eq:3}). As we can see the intensities of these sources
depend on the electron density and temperature of the low velocity
component.  With the derived [N\,{\sc ii}] 122$\mu$m and 205$\mu$m
intensities of the low velocity component we can obtain
\begin{equation}\label{eq:15}
   I^{high}_{122\mu m}= I_{122\mu m}-I^{low}_{122\mu m}
\end{equation}
and
\begin{equation}\label{eq:16}
   I^{high}_{205\mu m}= I_{205\mu m}-I^{low}_{205\mu m},
\end{equation}
and the electron density for the high velocity component is derived
from the [N\,{\sc ii}] 122$\mu$m/205$\mu$m ratio and an electron
temperature. With the intensity of either the [N\,{\sc ii}] 122$\mu$m
or 205$\mu$m lines, the electron temperature, density, and the
measured EM from the RRL, we can derive the ionized Nitrogen abundance
using Equation~(\ref{eq:1}),(\ref{eq:2}), and (\ref{eq:6}).

Note however, that the electron temperatures and densities of each LSR
velocity component cannot be independently derived.  This is because
both [N\,{\sc ii}] 122$\mu$m and 205$\mu$m and radio continuum
intensities, which are used to determine the electron density and
temperature, respectively, are velocity unresolved and thus correspond
to the sum of the intensities arising from each velocity component.
To investigate the range of possible solutions for the N$^+$/H$^+$
ratio for the high velocity component given these uncertainties, we
evaluated this quantity using the method described above, for ranges
in the electron temperature, for the low and high LSR velocity
components, from 1000\,K to 25000\,K, of the [N\,{\sc ii}]
122$\mu$m/205$\mu$m ratio, corresponding to $n_e$=10--1000\,cm$^{-3}$
Additionally, we assumed a range for the N$^+$/H$^+$ ratio for the low
LSR velocity component that corresponds to the typical standard
deviation for $R_{\rm gal}>4$\,kpc of a factor of 2. Using 10$^6$
combinations of these parameters, we solved for the N$^+$/H$^+$
abundance ratio of the high velocity component.  These solutions were
constrained using the measured values of the continuum temperature and
the [N\,{\sc ii}] 205$\mu$m and 122$\mu$m lines. We used
Equation~(\ref{eq:8}) to, given the known RRL intensity of each
component, estimate the continuum brightness temperature of each
component that would result from a given temperature, with the
constraints that the continuum brightness temperature of each source
has a SNR above 5 and that the sum of the resulting continuum
brightness temperatures of each component is equal to the measured
value within its uncertainties.  Additionally, for a given [N\,{\sc
    ii}] 122$\mu$m/205$\mu$m ratio and temperature for the low LSR
component, we determined the corresponding electron density. With the
electron density, electron temperature, and the N$^+$/H$^+$ ratio for
this velocity component, we evaluated the [N\,{\sc ii}] 205$\mu$m and
122$\mu$m line intensities using Equations~(\ref{eq:11}) and
(\ref{eq:14}). We then used Equations~(\ref{eq:15}) and (\ref{eq:16})
to evaluate the corresponding [N\,{\sc ii}] intensities and electron
density of the high LSR velocity component, for a given electron
temperature for this velocity component.  The solutions were
constrained requiring that the derived [N\,{\sc ii}] intensities for
the high and low LSR velocity components have each a SNR above 5, and
that the sum of the [N\,{\sc ii}] lines from these components match
the observed values within their uncertainties.  In
Tables~\ref{tab:results1} and \ref{tab:results2} we list the average
value of all possible solutions for N$^+$/H$^+$ and the constrained
values of the electron temperature, electron density, and N$^+$ and
H$^+$ column densities, for the high LSR velocity component together
with their corresponding standard deviation.  In
Figure~\ref{fig:solution_sample}, we show two examples of the derived
N$^+$/H$^+$ and constrained parameters, showing that this approach
results in well constrained parameters.

\begin{figure*}[t]
\centering
\includegraphics[angle=0,width=0.85\textwidth,angle=0]{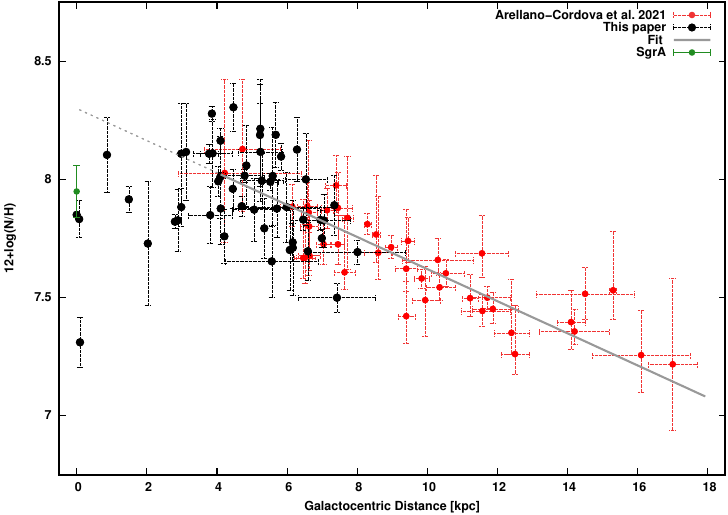}
\caption{The distribution of Nitrogen abundances as a function of
  Galactocentric distance. We show our sample ranging from 0 to 8\,kpc
  in the inner galaxy. We also include data points presented by
  \citet{Arellano-Cordova2021} sampling the Solar neighborhood and
  Outer galaxy from 6 to 17\,kpc. These datasets together represent a
  continuous sample of the Nitrogen abundance over the disk of the
  Milky Way from 0 to 17\,kpc.  We also show the ionized Nitrogen
  abundance derived in Sagittarius A and discussed in
  Section~\ref{sec:nitr-abund-determ} in dark green.
}\label{fig:Nitrogen_abundance}
\end{figure*}

  \begin{figure*}[t]
\centering
\includegraphics[angle=0,width=\textwidth,angle=0]{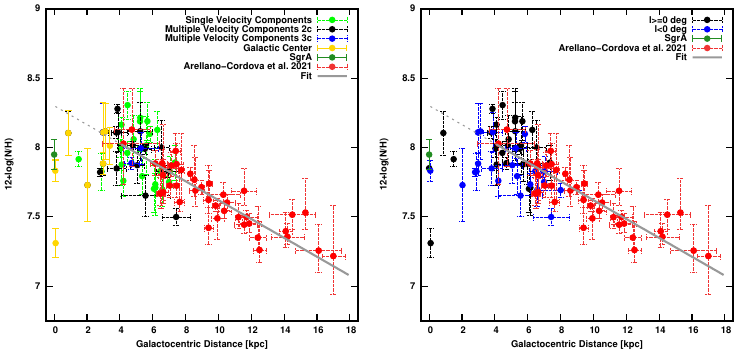}
\caption{({\it left}) Ionized Nitrogen abundances as a function of
  Galactocentric distance separated in the different cases in which
  they were derived, with single velocity components shown in green,
  double components in black, triple components in green, and LOS
  associated with the central molecular zone in shown in yellow.  We
  also show the ionized Nitrogen abundance derived in Sgr A and
  discussed in Section~\ref{sec:nitr-abund-determ} in dark
  green. ({\it right)} Ionized Nitrogen abundances as a function of
  Galactocentric distance separated between those derived from LOS
  with $l \geq 0$\degr\ (black) and
  $l<0$\degr\ (blue). }\label{fig:abudances2}
\end{figure*}

\begin{figure*}[t]
\centering
\includegraphics[angle=0,width=\textwidth,angle=0]{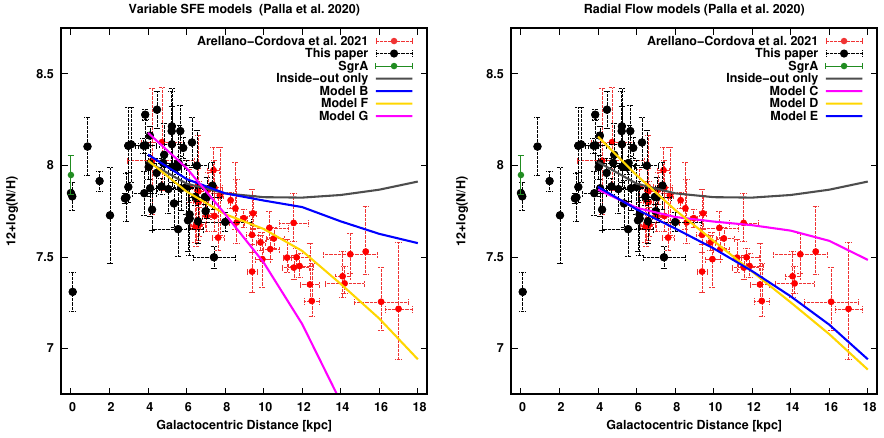}
\caption{Comparison between the observed Nitrogen abundance
  distribution as a function of Galactocentric distance and model
  calculations presented by \citet{Palla2020} for $R_{\rm
    gal}>4$\,kpc. These models include a prediction of inside-out
  growth only (dark grey), as well as models with variable star
  formation efficiency (SFE; {\it left panel}) and radial flows ({\it
    right panel}). Note that the absolute Nitrogen abundance values in
  the \citet{Palla2020} models were divided by a factor of 2.6 to
  match our observations.}
\label{fig:palla}
\end{figure*}

\begin{deluxetable*}{lrrcccccc} 
\tabletypesize{\footnotesize} \centering \tablecolumns{8} \small
\tablewidth{0pt} \tablecaption{Derived Nitrogen Abundances in the Galactic Plane for $l\geq 0$\degr } \tablenum{7}
\tablehead{\colhead{LOS} & \colhead{$l$} & \colhead{$b$} & \colhead{$n_e$} & \colhead{$T_e$} & \colhead{$N({\rm N}^+)$} & \colhead{$N({\rm H}^+)$} & \colhead{12+log(${\rm N}^+/{\rm H}^+$)}  & \colhead{$R_{\rm gal}$}  \\ 
\colhead{}   & \colhead{[\degr]}   & \colhead{[\degr]}   & \colhead{[cm$^{-3}$]} & \colhead{[K]} &   \colhead{[$10^{17}{\rm cm}^{-2}$]} & \colhead{[$10^{20}{\rm cm}^{-2}$]} &   \colhead{}  & \colhead{[kpc]} \\}
\startdata
G000.0+0.0 & 0.000  & 0.0  &  121.5 $\pm$ 0.5 & 11910  $\pm$ 97 & 35.8  $\pm$ 0.12 & 50.6  $\pm$ 0.9  & 7.8  $\pm$ 0.0  & 0.0   $\pm$  0.0  \\
G000.5+0.0 & 0.500  & 0.0  &  40.9 $\pm$ 23.4 & 12855  $\pm$ 706 & 9.9  $\pm$ 3.17 & 49.7  $\pm$ 19.3  & 7.3  $\pm$ 0.1  & 0.1   $\pm$  0.0  \\
G003.5+0.0 & 3.478  & 0.0  &  16.3 $\pm$ 0.3 & 5421  $\pm$ 1340 & 3.9  $\pm$ 0.04 & 2.4  $\pm$ 1.1  & 8.2  $\pm$ 0.2  & 5.2   $\pm$  0.0  \\
G004.3+0.0 & 4.348  & 0.0  &  24.7 $\pm$ 0.8 & 7791  $\pm$ 3330 & 2.2  $\pm$ 0.04 & 1.6  $\pm$ 1.4  & 8.1  $\pm$ 0.4  & 5.2   $\pm$  0.0  \\
G005.2+0.0 & 5.217  & 0.0  &  6.6 $\pm$ 0.5 & 7513  $\pm$ 1844 & 11.7  $\pm$ 0.29 & 7.6  $\pm$ 3.8  & 8.2  $\pm$ 0.2  & 5.2   $\pm$  0.0  \\
G006.1+0.0 & 6.087  & 0.0  &  18.8 $\pm$ 7.2 & 6969  $\pm$ 1992 & 5.3  $\pm$ 1.21 & 4.2  $\pm$ 1.5  & 8.1  $\pm$ 0.2  & 0.9   $\pm$  0.0  \\
G007.8+0.0 & 7.826  & 0.0  &  27.5 $\pm$ 0.5 & 10043  $\pm$ 1644 & 3.4  $\pm$ 0.03 & 4.5  $\pm$ 1.5  & 7.9  $\pm$ 0.1  & 6.0   $\pm$  1.4  \\
G008.7+0.0 & 8.696  & 0.0  &  14.3 $\pm$ 0.8 & 12852  $\pm$ 3316 & 11.1  $\pm$ 0.25 & 21.7  $\pm$ 12.0  & 7.7  $\pm$ 0.2  & 5.5   $\pm$  0.0  \\
G010.4+0.0 & 10.435  & 0.0  &  24.3 $\pm$ 0.5 & 6129  $\pm$ 367 & 10.1  $\pm$ 0.13 & 12.3  $\pm$ 1.5  & 7.9  $\pm$ 0.1  & 1.5   $\pm$  0.0  \\
G011.3+0.0 & 11.304  & 0.0  &  8.3 $\pm$ 0.6 & 4556  $\pm$ 707 & 10.5  $\pm$ 0.31 & 6.8  $\pm$ 2.2  & 8.2  $\pm$ 0.1  & 5.7   $\pm$  0.0  \\
G012.2+0.0 & 12.174  & 0.0  &  23.4 $\pm$ 0.4 & 5800  $\pm$ 648 & 10.9  $\pm$ 0.12 & 5.4  $\pm$ 1.2  & 8.3  $\pm$ 0.1  & 4.5   $\pm$  0.0  \\
G013.0+0.0 & 13.043  & 0.0  &  29.6 $\pm$ 0.4 & 7647  $\pm$ 1046 & 3.8  $\pm$ 0.03 & 5.1  $\pm$ 1.4  & 7.9  $\pm$ 0.1  & 5.7   $\pm$  1.2  \\
G013.9+0.0 & 13.913  & 0.0  &  17.9 $\pm$ 0.6 & 3501  $\pm$ 170 & 7.2  $\pm$ 0.15 & 9.4  $\pm$ 0.9  & 7.9  $\pm$ 0.0  & 4.7   $\pm$  0.0  \\
G014.8+0.0 & 14.783  & 0.0  &  21.6 $\pm$ 0.7 & 10169  $\pm$ 1904 & 7.2  $\pm$ 0.13 & 6.3  $\pm$ 2.5  & 8.1  $\pm$ 0.2  & 4.8   $\pm$  0.0  \\
G016.5+0.0 & 16.522  & 0.0  &  25.9 $\pm$ 0.2 & 5716  $\pm$ 1301 & 4.5  $\pm$ 0.02 & 3.4  $\pm$ 1.6  & 8.1  $\pm$ 0.2  & 5.2   $\pm$  0.6  \\
G020.0+0.0 & 20.000  & 0.0  &  12.5 $\pm$ 0.2 & 6479  $\pm$ 1706 & 4.7  $\pm$ 0.03 & 6.9  $\pm$ 3.8  & 7.8  $\pm$ 0.2  & 3.9   $\pm$  0.6  \\
G020.9+0.0 & 20.870  & 0.0  &  25.0 $\pm$ 0.7 & 6335  $\pm$ 566 & 6.5  $\pm$ 0.11 & 7.2  $\pm$ 1.3  & 8.0  $\pm$ 0.1  & 4.5   $\pm$  0.0  \\
G021.7+0.0 & 21.739  & 0.0  &  29.2 $\pm$ 0.7 & 12850  $\pm$ 1891 & 9.2  $\pm$ 0.13 & 6.9  $\pm$ 2.2  & 8.1  $\pm$ 0.1  & 6.3   $\pm$  0.0  \\
G023.5+0.0 & 23.478  & 0.0  &  26.0 $\pm$ 0.2 & 5949  $\pm$ 339 & 17.7  $\pm$ 0.11 & 12.1  $\pm$ 1.4  & 8.2  $\pm$ 0.1  & 4.1   $\pm$  0.0  \\
G024.3+0.0 & 24.348  & 0.0  &  18.1 $\pm$ 0.3 & 5744  $\pm$ 276 & 14.5  $\pm$ 0.13 & 11.3  $\pm$ 1.1  & 8.1  $\pm$ 0.0  & 3.8   $\pm$  0.3  \\
G025.2+0.0 & 25.217  & 0.0  &  12.8 $\pm$ 0.4 & 9093  $\pm$ 3292 & 14.9  $\pm$ 0.18 & 11.9  $\pm$ 9.2  & 8.1  $\pm$ 0.3  & 4.4   $\pm$  0.6  \\
G026.1+0.0 & 26.087  & 0.0  &  38.8 $\pm$ 0.4 & 5630  $\pm$ 198 & 14.9  $\pm$ 0.12 & 7.8  $\pm$ 0.5  & 8.3  $\pm$ 0.0  & 3.9   $\pm$  0.0  \\
G027.0+0.0 & 26.956  & 0.0  &  22.0 $\pm$ 0.3 & 10085  $\pm$ 1915 & 6.7  $\pm$ 0.04 & 6.6  $\pm$ 2.7  & 8.0  $\pm$ 0.2  & 4.1   $\pm$  0.0  \\
G028.7+0.0 & 28.696  & 0.0  &  37.6 $\pm$ 0.6 & 8663  $\pm$ 612 & 10.6  $\pm$ 0.11 & 10.9  $\pm$ 1.6  & 8.0  $\pm$ 0.1  & 4.0   $\pm$  0.0  \\
G030.0+0.0 & 30.000  & 0.0  &  31.8 $\pm$ 0.9 & 13227  $\pm$ 2184 & 8.6  $\pm$ 0.14 & 11.5  $\pm$ 4.0  & 7.9  $\pm$ 0.2  & 4.1   $\pm$  0.0  \\
G031.3+0.0 & 31.277  & 0.0  &  25.8 $\pm$ 0.4 & 5548  $\pm$ 170 & 12.1  $\pm$ 0.12 & 11.7  $\pm$ 0.7  & 8.0  $\pm$ 0.0  & 4.8   $\pm$  0.6  \\
G036.4+0.0 & 36.383  & 0.0  &  20.8 $\pm$ 1.7 & 7445  $\pm$ 1652 & 3.3  $\pm$ 0.16 & 3.2  $\pm$ 1.5  & 8.0  $\pm$ 0.2  & 5.6   $\pm$  0.0  \\
G037.7+0.0 & 37.660  & 0.0  &  17.1 $\pm$ 0.5 & 5299  $\pm$ 214 & 9.6  $\pm$ 0.16 & 9.8  $\pm$ 0.8  & 8.0  $\pm$ 0.0  & 5.5   $\pm$  0.2  \\
G041.5+0.0 & 41.489  & 0.0  &  46.8 $\pm$ 0.5 & 10619  $\pm$ 1481 & 2.2  $\pm$ 0.02 & 2.8  $\pm$ 0.8  & 7.9  $\pm$ 0.1  & 7.3   $\pm$  0.0  \\
G044.0+0.0 & 44.043  & 0.0  &  13.8 $\pm$ 0.3 & 4308  $\pm$ 929 & 3.4  $\pm$ 0.05 & 3.4  $\pm$ 1.5  & 8.0  $\pm$ 0.2  & 6.5   $\pm$  0.6  \\
G045.3+0.0 & 45.319  & 0.0  &  29.3 $\pm$ 1.1 & 6955  $\pm$ 1313 & 0.9  $\pm$ 0.02 & 1.8  $\pm$ 0.7  & 7.7  $\pm$ 0.2  & 6.1   $\pm$  0.0  \\
G049.1+0.0 & 49.149  & 0.0  &  21.7 $\pm$ 1.7 & 13245  $\pm$ 2858 & 3.3  $\pm$ 0.13 & 6.5  $\pm$ 3.0  & 7.7  $\pm$ 0.2  & 6.2   $\pm$  0.0  \\
G054.3+0.0 & 54.255  & 0.0  &  46.0 $\pm$ 1.0 & 10888  $\pm$ 2869 & 1.1  $\pm$ 0.02 & 1.6  $\pm$ 0.9  & 7.8  $\pm$ 0.2  & 6.6   $\pm$  0.0  \\
\enddata
\tablenotetext{}{}
\label{tab:results1}
\end{deluxetable*}

\begin{deluxetable*}{lrrcccccc} 
\tabletypesize{\footnotesize} \centering \tablecolumns{8} \small
\tablewidth{0pt} \tablecaption{Derived Nitrogen Abundances in the Galactic Plane for $l<0$\degr} \tablenum{8}
\tablehead{\colhead{LOS} & \colhead{$l$} & \colhead{$b$} & \colhead{$n_e$} & \colhead{$T_e$} & \colhead{$N({\rm N}^+)$} & \colhead{$N({\rm H}^+)$} & \colhead{12+log(${\rm N}^+/{\rm H}^+$)}  & \colhead{$R_{\rm gal}$}  \\ 
\colhead{}   & \colhead{[\degr]}   & \colhead{[\degr]}   & \colhead{[cm$^{-3}$]} & \colhead{[K]} &   \colhead{[$10^{17}{\rm cm}^{-2}$]} & \colhead{[$10^{20}{\rm cm}^{-2}$]} &   \colhead{}  & \colhead{[kpc]} \\}
\startdata
G302.6+0.0 & 302.553  & 0.0  &  12.5 $\pm$ 0.2 & 4015  $\pm$ 481 & 5.3  $\pm$ 0.04 & 7.9  $\pm$ 2.0  & 7.8  $\pm$ 0.1  & 7.1   $\pm$  0.0  \\
G305.1+0.0 & 305.106  & 0.0  &  4.9 $\pm$ 0.2 & 4321  $\pm$ 162 & 31.5  $\pm$ 0.34 & 55.9  $\pm$ 4.9  & 7.8  $\pm$ 0.0  & 7.0   $\pm$  0.0  \\
G306.4+0.0 & 306.383  & 0.0  &  9.9 $\pm$ 0.4 & 8090  $\pm$ 1043 & 3.9  $\pm$ 0.06 & 7.8  $\pm$ 2.2  & 7.7  $\pm$ 0.1  & 6.6   $\pm$  0.0  \\
G307.7+0.0 & 307.660  & 0.0  &  15.4 $\pm$ 0.9 & 8658  $\pm$ 309 & 5.9  $\pm$ 0.17 & 8.7  $\pm$ 0.9  & 7.8  $\pm$ 0.0  & 6.5   $\pm$  0.0  \\
G310.2+0.0 & 310.213  & 0.0  &  62.3 $\pm$ 2.1 & 8792  $\pm$ 539 & 0.6  $\pm$ 0.02 & 1.9  $\pm$ 0.3  & 7.5  $\pm$ 0.1  & 7.4   $\pm$  1.1  \\
G314.0+0.0 & 314.043  & 0.0  &  26.4 $\pm$ 0.3 & 8349  $\pm$ 706 & 2.5  $\pm$ 0.02 & 4.6  $\pm$ 0.8  & 7.7  $\pm$ 0.1  & 6.2   $\pm$  0.0  \\
G316.6+0.0 & 316.596  & 0.0  &  21.0 $\pm$ 0.6 & 8844  $\pm$ 372 & 9.8  $\pm$ 0.16 & 14.5  $\pm$ 1.4  & 7.8  $\pm$ 0.0  & 7.0   $\pm$  0.8  \\
G317.9+0.0 & 317.872  & 0.0  &  25.3 $\pm$ 0.9 & 9234  $\pm$ 477 & 5.9  $\pm$ 0.12 & 11.9  $\pm$ 1.4  & 7.7  $\pm$ 0.1  & 8.0   $\pm$  1.4  \\
G326.8+0.0 & 326.808  & 0.0  &  26.5 $\pm$ 0.4 & 5199  $\pm$ 317 & 11.2  $\pm$ 0.12 & 9.0  $\pm$ 1.2  & 8.1  $\pm$ 0.1  & 5.8   $\pm$  0.0  \\
G330.0+0.0 & 330.000  & 0.0  &  39.0 $\pm$ 1.0 & 13446  $\pm$ 6062 & 1.6  $\pm$ 0.02 & 4.2  $\pm$ 4.1  & 7.6  $\pm$ 0.4  & 5.2   $\pm$  0.5  \\
G331.7+0.0 & 331.739  & 0.0  &  29.5 $\pm$ 1.0 & 16634  $\pm$ 4501 & 7.6  $\pm$ 0.14 & 9.2  $\pm$ 5.3  & 7.9  $\pm$ 0.2  & 5.7   $\pm$  0.0  \\
G332.6+0.0 & 332.609  & 0.0  &  25.6 $\pm$ 0.9 & 3941  $\pm$ 571 & 4.8  $\pm$ 0.11 & 6.5  $\pm$ 2.0  & 7.9  $\pm$ 0.1  & 5.1   $\pm$  0.6  \\
G333.5+0.0 & 333.478  & 0.0  &  36.8 $\pm$ 0.6 & 7703  $\pm$ 991 & 8.7  $\pm$ 0.09 & 8.8  $\pm$ 2.4  & 8.0  $\pm$ 0.1  & 5.3   $\pm$  0.6  \\
G336.1+0.0 & 336.087  & 0.0  &  28.5 $\pm$ 0.3 & 8161  $\pm$ 1282 & 18.5  $\pm$ 0.11 & 14.3  $\pm$ 4.7  & 8.1  $\pm$ 0.1  & 3.9   $\pm$  0.6  \\
G337.0+0.0 & 336.957  & 0.0  &  31.9 $\pm$ 0.3 & 15478  $\pm$ 2066 & 23.0  $\pm$ 0.14 & 32.6  $\pm$ 9.0  & 7.8  $\pm$ 0.1  & 3.8   $\pm$  0.6  \\
G337.8+0.0 & 337.826  & 0.0  &  22.6 $\pm$ 10.8 & 14731  $\pm$ 5334 & 16.7  $\pm$ 5.64 & 13.3  $\pm$ 6.2  & 8.1  $\pm$ 0.2  & 3.1   $\pm$  0.0  \\
G338.7+0.0 & 338.696  & 0.0  &  28.0 $\pm$ 0.3 & 6405  $\pm$ 906 & 3.8  $\pm$ 0.03 & 6.2  $\pm$ 1.8  & 7.8  $\pm$ 0.1  & 5.4   $\pm$  0.0  \\
G342.2+0.0 & 342.174  & 0.0  &  11.0 $\pm$ 0.6 & 8643  $\pm$ 1176 & 11.5  $\pm$ 1.98 & 15.1  $\pm$ 2.9  & 7.9  $\pm$ 0.1  & 3.0   $\pm$  0.0  \\
G343.9+0.0 & 343.913  & 0.0  &  21.7 $\pm$ 11.1 & 16864  $\pm$ 5699 & 4.3  $\pm$ 2.09 & 3.6  $\pm$ 2.3  & 8.1  $\pm$ 0.2  & 3.0   $\pm$  0.0  \\
G345.7+0.0 & 345.652  & 0.0  &  25.2 $\pm$ 10.2 & 15869  $\pm$ 6014 & 10.5  $\pm$ 2.68 & 21.7  $\pm$ 12.5  & 7.7  $\pm$ 0.3  & 2.0   $\pm$  0.0  \\
G346.5+0.0 & 346.522  & 0.0  &  30.4 $\pm$ 1.5 & 8642  $\pm$ 1394 & 3.4  $\pm$ 0.11 & 7.5  $\pm$ 2.6  & 7.7  $\pm$ 0.1  & 5.6   $\pm$  1.3  \\
G349.1+0.0 & 349.130  & 0.0  &  52.2 $\pm$ 0.5 & 5674  $\pm$ 207 & 13.1  $\pm$ 0.09 & 19.7  $\pm$ 1.3  & 7.8  $\pm$ 0.0  & 2.8   $\pm$  0.0  \\
G350.9+0.0 & 350.870  & 0.0  &  14.4 $\pm$ 1.3 & 5845  $\pm$ 1572 & 3.1  $\pm$ 0.14 & 6.5  $\pm$ 3.8  & 7.7  $\pm$ 0.2  & 4.4   $\pm$  0.0  \\
G353.5+0.0 & 353.478  & 0.0  &  17.0 $\pm$ 5.1 & 12417  $\pm$ 3271 & 7.0  $\pm$ 1.61 & 6.9  $\pm$ 2.2  & 8.0  $\pm$ 0.1  & 3.4   $\pm$  0.0  \\
G354.3+0.0 & 354.348  & 0.0  &  31.8 $\pm$ 2.1 & 8488  $\pm$ 1212 & 2.7  $\pm$ 0.11 & 4.0  $\pm$ 1.2  & 7.8  $\pm$ 0.1  & 2.9   $\pm$  0.0  \\
G355.2+0.0 & 355.217  & 0.0  &  87.8 $\pm$ 3.0 & 12528  $\pm$ 1706 & 2.7  $\pm$ 0.07 & 4.7  $\pm$ 1.2  & 7.8  $\pm$ 0.1  & 4.2   $\pm$  0.0  \\
G356.1+0.0 & 356.087  & 0.0  &  35.3 $\pm$ 0.3 & 10465  $\pm$ 2529 & 2.8  $\pm$ 0.01 & 3.1  $\pm$ 1.5  & 8.0  $\pm$ 0.2  & 4.2   $\pm$  0.0  \\
G359.5+0.0 & 359.500  & 0.0  &  47.3 $\pm$ 21.4 & 20202  $\pm$ 2209 & 13.5  $\pm$ 3.96 & 19.9  $\pm$ 6.2  & 7.8  $\pm$ 0.1  & 0.1   $\pm$  0.0  \\
\enddata
\tablenotetext{}{}
\label{tab:results2}
\end{deluxetable*}

\subsection{The distribution of Nitrogen abundances in the disk of the Milky Way.}
\label{sec:distr-nitr-abund}

In Figure~\ref{fig:Nitrogen_abundance}, we show the distribution of
Nitrogen abundances as a function of Galactocentric distance derived
from our sample in the range from 0 to 8\,kpc in the inner Galaxy. We
used only data with RRL emission above a SNR of 5, and a radio
continuum brightness temperature used to determine the electron
temperature with SNR above 10. These criteria result in a sample of 41
positions for which we consider the data to be of high quality.  We
also show, in dark green, the Nitrogen abundance derived in Sgr A and
discussed in Section~\ref{sec:nitr-abund-determ}.  We also include a
sample of Nitrogen abundances derived in 42 H\,{\sc ii} regions
presented by \citet{Arellano-Cordova2021}, using optical spectral
lines, sampling the Galactic plane from 4 to 17\,kpc.  We find that
the Nitrogen abundances derived here and those those derived with
optical spectral line observations are in excellent agreement in the
Galactocentric distance region they overlap. Taken together, these
data sets represent a continuous sample of the Nitrogen abundance over
the disk of the Milky Way from 0 to 17\,kpc.

Both our Nitrogen abundances and those from
\citet{Arellano-Cordova2021} are determined in low ionization regions
that have a negligible fraction of doubly ionized Nitrogen. Therefore
we can assume for both data sets that N/H$\approx$N$^+$/H$^+$. The
agreement between our Nitrogen abundances and those derived using
optical data is in contrast to discrepancies between FIR and optical
derived abundances and the N/O ratio as reported by \citet{Rudolph2006}
and \citet{Arellano-Cordova2021}.   \citet{Arellano-Cordova2021}
  compared the optical based observations of the N/H abundance
  gradient with those derived using mid-- and far--infrared
  observations by \citet{Rudolph2006} finding that the latter show a
  significantly steeper gradient and larger dispersion, compared to
  those derived with optical lines. A possible explanation for this
  discrepancy is that \citet{Rudolph2006} abundance determination is
  based on observations of spectral lines tracing high ionization
  states, such as [N\,{\sc iii}] and [O\,{\sc iii}] lines, and
  therefore they rely in a correction factor for lower ionization gas
  that can introduce significant uncertainties in the measurements.

In the left panel of Figure~\ref{fig:abudances2} we show the ionized
Nitrogen abundances as a function of Galactocentric distance separated
into the different cases in which they were derived, with single
velocity components shown in red, double components in black, triple
components in green, and LOS associated with the central molecular
zone shown in yellow. In the right panel of
Figure~\ref{fig:abudances2} we show the ionized Nitrogen abundance as
a function of Galactocentric distance separated from those derived for
LOS with $l \geq 0$\degr\ (black) and $l<0$\degr\ (blue). We find that
the ionized Nitrogen abundances for LOS with $l \geq 0$\degr\ are on
average 40\% larger than for $l<0$\degr. This difference coincides
with a similar asymmetry in the star formation rate distribution in the
Milky Way \citep{Elia2022}, suggesting that metal production is
enhanced in the region with $l \geq 0$\degr\ compared to that for
$l<0$\degr.

We find that the Nitrogen abundance in the Milky Way decreases from
about 4\,kpc out to 17\,kpc, while having a flat distribution from
4\,kpc to the Galactic center. Observations of Cepheids and red giants
also show that different elements, including iron (Fe), have
abundances close to the Galactic center that are lower than predicted
by extrapolating the abundance distribution at larger radii
\citep{Davies2009,Najarro2009,Hayden2014,Martin2015,Andrievsky2016},
in agreement with our results.  We did not attempt to fit our data in
the 4\,kpc$<R_{\rm gal}<8$\,kpc range, as this range is too narrow for
enabling an accurate representation of N/H across the Milky Way.
Instead, we combined our data set with that from
\citet{Arellano-Cordova2021} to obtain a fit to the distribution of
Nitrogen abundance in the Galactic plane between 4\,kpc and 17\,kpc.
We used the orthogonal bi-variate error and intrinsic scatter method
\citep[BES][]{Akritas1996}, including a bootstrap resampling error
analysis, resulting in,
\begin{equation}\label{eq:12}
  12+\log({\rm N}^+/{\rm H}^{+})= 8.30\pm0.04-(0.068\pm0.005)R_{\rm gal}.
\end{equation}
The slope of our fit is consistent within its uncertainties to that
derived by \citet{Arellano-Cordova2021}, $-$0.057\,dex\,kpc$^{-1}$,
using optical lines, but is shallower than that derived by
\citet{Pineda2019}, $-$0.076\,dex\,kpc$^{-1}$, using the same method
presented here in a smaller sample of 11 LOS, and that derived by
\citet{Rudolph2006}, $-$0.085\,dex\,kpc$^{-1}$. The slope of our fit
is steeper than that derived for O/H by \citet{Arellano-Cordova2021},
$-0.042$\,dex\,kpc$^{-1}$. This difference can be understood as a
larger number of older intermediate mass stars, that can contribute
additional Nitrogen to the ISM, are present in the inner Galaxy.  The
typical dispersion from the fit for our data set for $R_{\rm
  gal}>4$\,kpc is 0.16\,dex and 0.14\,dex when also considering the
data set from \citet{Arellano-Cordova2021}. These dispersions are
somewhat larger than those reported by \citet{Arellano-Cordova2021},
of 0.1\,dex, but are consistent with the suggestion from this work
that azimuthal variations are not significant for Nitrogen.

Note that in \citet{Pineda2019} we assumed electron temperatures from
\citet{Balser2011} instead of deriving them using radio continuum data
as is done here. The \cite{Balser2011} electron temperature gradient
was derived from H\,{\sc ii} regions between 5\,kpc and 8\,kpc, but we
extrapolated the fit down to the Galactic center. Data presented by
\citet{Quireza2006} show electron temperatures in the Galactic center
that are significantly larger than predicted by extrapolating the fit
from electron temperatures at larger Galactocentric distances inward
to $R_{\rm gal}$=0\,kpc. This underestimation of the electron
temperature at the Galactic center resulted in an overestimation of
the Nitrogen abundances in this region as found by \citet{Pineda2019}.

\subsubsection{Comparison with chemical evolution models of the Milky Way.}
\label{sec:comp-with-chem}

Metallicity gradients in the disk of Galaxies are formed when star
formation is more efficient in their inner parts compared with their
outer parts \citep{Matteucci2021}.  Such a gradient in the star
formation efficiency can be produced by ``inside--out'' galaxy
formation, in which the disk of galaxies form by gas accretion with a
rate that is faster in the inner Galaxy compared with the outer Galaxy
\citep{Larson1976,Matteucci1989,Boissier1999,Pilkington2012}. In these
models, the measured slope of the gradient constrains the galaxy
accretion rate.  However, other mechanisms can also predict and/or
steepen a metallicity gradient such as the presence of a density
threshold for star formation, a star formation efficiency that
decreases with Galactocentric distances, and inwards radial flows
\citep{Kubryk2015,Palla2020,Grisoni2021}.  \citet{Palla2020} presented
set of chemical evolution models of the Milky Way that assume a
two--infall model \citep{Chiappini1997,Romano2010}, in which the thick
and thin disks were formed in two accretion episodes separated by
$\sim$3.25\,Gyr, to study the abundance distribution of several
elements as a function of Galactocentric distance.  These models that
include the effects of radially variable star formation efficiency
(SFE) and radial flows in addition to inside--out growth in the
determination of the radial distribution of element abundances in the
Milky Way.  The mechanism by which a variable SFE induces and/or
steepen an abundance gradient is that in the innermost regions of
galaxies the star formation rate is enhanced compared to the outer
regions, leading to an increased chemical enrichment.  Additionally,
radial flows can contribute to this effect. As gas moves toward the
inner parts of the galaxy, the star formation rate increases,
resulting in more significant metal production closer to the center
compared to the outer areas.

  In Figure~\ref{fig:palla} we compare the observed Nitrogen abundance
  distribution with the models presented by \citet{Palla2020} for
  Nitrogen in the Milky Way. In both panels of Figure~\ref{fig:palla}
  we show a model that includes inside--out growth only (dark grey),
  which shows a shallower distribution compared with observations,
  suggesting that this mechanism alone is insufficient to reproduce
  the observed Nitrogen abundance gradient. In the left panel
  Figure~\ref{fig:palla} we show models with variable star formation
  efficiency labeled B, F, and G, and the right panel of
  Figure~\ref{fig:palla} we show models with radial flows labeled C,
  D, and E in \citet{Palla2020}. We refer to \citet{Palla2020} for
  specific parameters assumed for these models. As we can see, in
  agreement with the conclusion in \citet{Palla2020}, the variable SFE
  model F, and the radial flow model E, agree the best with the
  observed Nitrogen abundance gradients. Thus, in a addition to
  inside-out growth, variable SFE and/or radial flows are necessary to
  explain the observed of the Nitrogen abundance gradient for $R_{\rm
    gal}>4$\,kpc.

Note that absolute value of the predicted Nitrogen abundances in the
\citet{Palla2020} models are on average a factor of 2.6 (0.41\,dex)
larger than the observed values. In Figure~\ref{fig:palla} we adjusted
the model predicted absolute Nitrogen abundances by this factor so
that their average value coincides with the Nitrogen abundance
predicted by the fit in Equation~(\ref{eq:12}) at 4\,kpc. The
\citet{Palla2020} models assume stellar yields for massive stars from
\citet{Kobayashi2006,Kobayashi2011}, and for low--intermediate mass
stars from \citet{Karakas2010}, and uncertainties in stellar yield
calculations might result in an overestimation in the production of
Nitrogen.  In addition, based on optical spectroscopy of stars that
show the presence two, higher and lower metallity, populations of
massive stars in the disk of the Milky Way, it has been recently
hypothetized that a recent accretion episode (in addition to the
two--infall model) of low metalliciy gas in the Milky Way disk in the
last ~2.7\,Gyr have resulted in a a general ISM metal impoverishment
\citep[][Palla et al. 2024 in prep.]{Spitoni2023}. Given that the
Nitrogen abundances derived here those by \citet{Arellano-Cordova2021}
trace the recent production of Nitrogen in the Milky way, our
observations would be consistent with this hypothesis.

In Figure~\ref{fig:Nitrogen_abundance}, we see a peak in the N/H
abundance at $R_{\rm gal}\simeq$4\,kpc, which is associated with the
outermost part of the Galactic Bar \citep{Benjamin2005}. A bar
potential can efficiently redistribute angular momentum and mass in
galaxies, and the radial flows produced by such a potential are
expected to mix elemental abundances, flattening any abundance
gradient over time \citep{Friedli1993,Friedli1994}.  The star
formation rate in the Milky Way peaks at $R_{\rm gal}\approx 4$\,kpc,
and do not continue to rise for smaller Galactocentric radii
\citep{Pineda2014,Elia2022}. Therefore the formation of new elements
in the inner 4\,kpc must be less efficient.  Note that the observed
reduction in the star formation rate in the inner Galaxy is likely due
to a reduction of the number of star forming regions per unit area
compared with larger Galactocentric distances.

Chemical evolution models predict elemental abundances to reach an
equilibrium value in regions where the production of new elements is
balanced by metal consumption by star formation and expulsion by
outflows \citep{Peng2014,Weinberg2017}.  The time scales for reaching
the equilibrium abundance are different for each element, with
elements such as Oxygen and Nitrogen reaching their equilibrium at the
timescale at which gas is being depleted by star formation and
outflows, while Fe reaches equilibrium at this timescale or that of
SNe Ia enrichment ($\sim$1.5\,Gyr), whichever is longer
\citep{Weinberg2017}. \citet{Belfiore2017} observed a flattening of
the O/H abundance in the center of galaxies with stellar masses
similar to that of the Milky Way, but observed that the O/N ratio
continues to rise toward the center of these galaxies. They
interpreted this result as Oxygen reaching an equilibrium abundance,
but Nitrogen is continuously being produced by secondary
nucleosynthetic production, in longer--living, intermediate mass
stars. We do not observe this effect in the center of the Milky Way,
suggesting that the timescales for a Nitrogen abundance equilibrium
has not been reached in the inner Milky Way. Note however, that the
star formation rate in the Milky Way peaks at $R_{\rm gal}=4$\,kpc
\citep{Pineda2014,Elia2022}, and thus the high star formation rate
required for the equilibrium hypothesis is not reached at smaller
radii.  We therefore favor radial flows induced by the stellar bar as
the most likely mechanism for the flattening of the Nitrogen abundance
in the innermost parts of the Galaxy.

\section{Summary}
\label{sec:conclusions}

 We presented a Galactic plane survey of Hydrogen Radio Recombination
 Lines (RRLs) observed with the NASA DSS--43 70m antenna and the Green
 Bank Telescope. We observed 108 lines--of--sights covering a range
 between $-135$\degr$< l < 60$\degr\ and $b=0$\degr\ in the Galactic
 plane. We combined these observations with observations of the
 [N\,{\sc ii}] 122$\mu$m and 205$\mu$m lines taken with the {\it
   Herschel} space observatory and [N\,{\sc iii}] 57$\mu$m taken with
 SOFIA/FIFI--LS, and radio continuum data, to characterize the
 distribution of the Nitrogen abundance across the disk of the Milky
 Way. In a sample of 41 LOS, where we have high enough
 signal--to--noise ratio, we studied the distribution of the ionized
 Nitrogen abundance relative to ionized Hydrogen covering
 galactocentric distances between 0 to 8\,kpc. Combined with existing
 determinations of the N/H abundance in the solar neighborhood and
 outer Galaxy, we are able to study for the first time the
 distribution of this quantity in the inner 16\,kpc of the Milky
 Way. The results of this work can be summarized as follows:

\begin{itemize}

 \item We find a Nitrogen abundance gradient extending over
   Galactocentric distances between 4 and 17\,kpc in  the Galactic
   plane, while for 0 to 4\,kpc we find a flat N/H distribution.

\item The gradient observed at Galactocentric distances larger that
  4\,kpc supports inside--out galaxy growth with the additional
  steepening resulting from variable star formation efficiency and/or
  radial flows in the Galactic disk.

\item The observed flattening of the Nitrogen abundance distribution
  in the inner 4\,kpc, which coincides with the start of the Galactic
  bar, can be associated with radial flows induced by the bar
  potential.

 \item We studied the ionization structure of a sub--sample of 8 LOS
   for which we obtained [N\,{\sc iii}] 57$\mu$m observations, and
   [O\,{\sc iii}] 88$\mu$m and 52$\mu$m observations in Sagittarius A.
   We find that most of the Nitrogen in our sample is likely singly
   ionized, which is consistent with their locations being in low
   ionization outskirts of H\,{\sc ii} regions, and that any highly
   ionized Nitrogen comes from compact high electron density H\,{\sc
     ii} regions.

  \end{itemize}

   Our observations demonstrate the power of using of far--infrared
   spectral lines and radio recombination lines, for an unobscured
   study of the ionization structure and the Nitrogen abundance
   distribution in galaxies. Far--infrared observatories, such as the
   GUSTO and ASTHROS balloons, and a future NASA far--infrared probe
   mission, together with ground based radio observatories, such as
   e.g. the GBT and NASA DSN antennas, can provide important insights
   in the chemical evolution in galaxies, which in turn provide
   important information to models of galaxy evolution.

\begin{acknowledgements}

 This research was carried out at the Jet Propulsion Laboratory,
 California Institute of Technology, under a contract with the
 National Aeronautics and Space Administration. The work in this
 publication was supported by NASA’s Astrophysics Data Analysis
 Program (ADAP) under grant No. 80NM0018F0610,18-2ADAP18-0196.
 We would like to acknowledge and express our gratitude to Javier
 Goicoechea for kindly providing the {\it Herschel}/PACS data in
 Sagittarius A. We also extend our sincere appreciation to Karla
 Arellano-Cordova for contributing the optical Nitrogen abundance data
 and to Marco Palla for supplying the model Nitrogen abundance
 distributions and for their insightful discussions. Their inputs and
 contributions have greatly enriched and strengthened our work. We
 also thank the anonymous referee for comments that significantly
 improved the paper.  This project made use of the Smithsonian
 Astrophysical Observatory $4 \times 32\mathrm{k}$-channel
 spectrometer (SAO32k) and the \emph{TAMS observatoryCtrl} observing
 system, which were developed by L.  Greenhill (Center for
 Astrophysics), I. Zaw (New York University Abu Dhabi), D. Price, and
 D. Shaff, with funding from SAO and the NYUAD Research Enhancement
 Fund and in-kind support from the Xilinx University Program. We thank
 West Virginia University for its financial support of GBT operations,
 which enabled the observations for this project. The National Radio
 Astronomy Observatory is a facility of the National Science
 Foundation operated under cooperative agreement by Associated
 Universities, Inc.  LDA and ML are supported by NSF grant AST1516021
 to LDA.  We thank the staff of the SOFIA Science Center for their
 help.  U.S. Government sponsorship acknowledged.

\end{acknowledgements}
\software{TMBIDL \citep{Bania2014}, GILDAS/CLASS \citep{Pety2005}}
\facilities{GBT, DSN/DSS--43, SOFIA, Herschel}
\appendix

\section{Dependence of Nitrogen abundance distribution on electron temperatures derived from different radio continuum surveys.}\label{sec:depend-nitr-abund}

\begin{figure*}[t]
\centering
\includegraphics[angle=0,width=1\textwidth,angle=0]{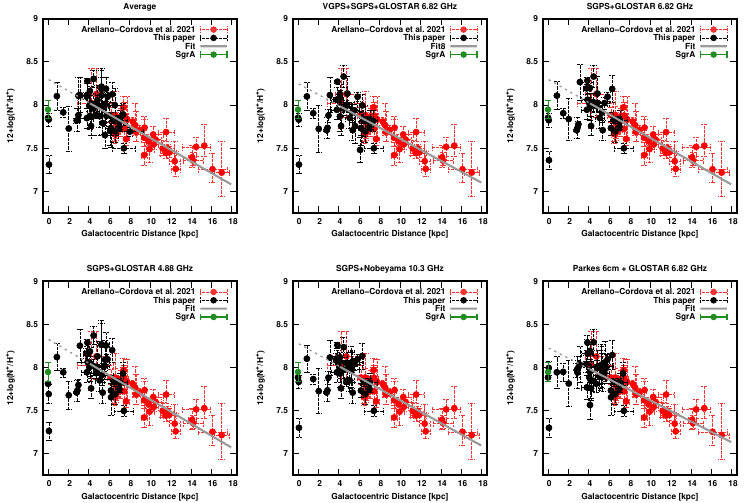}
\caption{The fractional abundance of ionized Nitrogen with respect to
  ionized Hydrogen as a function of Galactocentric distance derived
  for different combinations of radio continuum surveys. Note that the
  different surveys listed cover the majority of sources with $R_{\rm
    gal} > 5$\,kpc.}\label{fig:all_Xn}
\end{figure*}

In Figure~\ref{fig:all_Xn} we show the N$^{+}$/H$^{+}$ distribution as
a function of Galactocentric distance derived using electron
temperatures derived from each of the radio continuum surveys listed
in Table\,\ref{tab:radio_surveys}. In general, most LOS with
$l<0$\degr\ are covered by the SGPS and/or the Parkes 6\,cm
survey. For sources with $l\geq 0$\degr, all sources are covered by
either the Effelsberg/GLOSTAR survey at 4.88\,GHz and 6.82\,GHz and
the Nobeyama 10.3\,GHz survey. A subsample is also covered by the VGPS
survey at 1.4\,GHz. Each panel in Figure~\ref{fig:all_Xn} are labeled
with the combination of surveys used to cover the entire range of
Galactic longitudes. Note that some LOS do not appear in all panels
because they do not meet the SNR$>$10 criteria for the free--free
brightness temperature.  We also show the N$^{+}$/H$^{+}$ distribution
resulting from using the averaged free--free continuum temperature as
described in Section~\ref{sec:electr-temp-determ}. We find that
N$^{+}$/H$^{+}$ distribution is not significantly affected by the
choice of radio continuum survey used to derive electron temperatures.

\section{Comparison with Balser et al. 2015 electron temperatures. }
\label{sec:comp-with-citetb}
The uncertainties derived for the electron temperatures described in
Section~\ref{sec:electr-temp-determ} are based on the uncertainties of
the measurements of the continuum and RRL emission, and therefore do
not account for calibration uncertainties, that can vary for the
different frequency spectral bands used for the continuum emission,
and the accuracy of our correction for synchrotron emission. To assess
the sensitivity of the derived electron temperatures to these
uncertainties we compared our methodology to derive electron
temperatures against electron temperatures derived in a sample of
H\,{\sc ii} regions by \citet{Balser2015} in which calibration
uncertainties and synchrotron contribution are carefully assessed.

\citet{Balser2015} derived electron temperatures in a sample of 21
H\,{\sc ii} regions in the Galactic plane in both RRL and continuum
emission at 8.7\,GHz using the Green Bank Telescope. This sample was
selected to have a continuum intensity signal-to-noise ratio greater
than 10 and to be isolated enough so that spatial and spectral
blending can be avoided. Because the RRL and continuum observations
are done in the same band, with the same telescope, these measurements
do not suffer from calibration uncertainties. Also, because the
spatial structure in continuum is well isolated, the free--free and
synchrotron emission can be separated by fitting the compact
free--free emission with a Gaussian function and the more extended
synchrotron emission with a low order polynomial baseline.  Therefore
a comparison between the electron temperatures derived by
\citet{Balser2015} and those derived using our methodology can be used
to assess the impact of calibration uncertainties and
free--free/synchrotron separation in the electron temperatures
derived in our sample.

There are however a few differences between our sources and those used
by \citet{Balser2015}. Their sources are compact H\,{\sc ii} regions,
while our sources are associated with the outskirts of H\,{\sc ii}
regions. As discussed in Section~\ref{sec:electr-temp-determ}, the
beam filling effects in our Galactic plane sample are small but are
more significant for the \citet{Balser2015} sample. Also, the electron
densities in this sample are likely to be larger than the typical
densities in our sample. For this comparison, we will not apply a
non--LTE correction, as we did for our sample.

\begin{figure*}[t]
\centering
\includegraphics[angle=0,width=1\textwidth,angle=0]{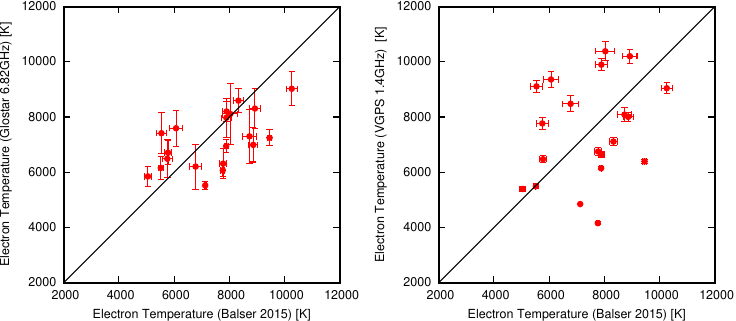}
\caption{({\it left }) Comparison between electron temperatures derived
  by \citet{Balser2015} using Green Bank Telescope and those derived
  with the same RRL emission but with continuum intensities derived
  from the GLOSTAR/Effelsberg 6.82\,GHz radio continuum survey.  ({\it
    right}) Comparison between electron temperatures derived by
  \citet{Balser2015} using Green Bank Telescope and those derived with
  the same RRL emission but with continuum intensities derived from
  the VGPS 1.4\,GHz radio continuum
  survey.}\label{fig:Balser_comparison}
\end{figure*}

In the left panel of Figure~\ref{fig:Balser_comparison}, we show a
comparison between the electron temperatures derived from RRL and
continuum intensities from \citet{Balser2015} and those derived using
our methodology with the RRL intensities from \citet{Balser2015} but
with the continuum intensities extracted from the Effelsberg 100\,m
GLOSTAR data at 6.82\,GHz. The angular resolution of the Effelsberg
data set is 106\arcsec\ which is somewhat larger than the
87\arcsec\ resolution of the GBT data. We find good agreement between
the electron temperatures using the two methods. The scatter from the
one--to--one correlation between the electron temperatures derived
from these two methods is on average 18\%. This difference can arise
from the different angular resolution of the data, which can be
important due to the compact H\,{\sc ii} emission in this sample, from
the uncertainties in the relative calibration between the Effelsberg
and GBT data sets, and/or from the separation between free-free and
synchrotron emission, with the latter estimated to be on average 19\%
of the total continuum emission at 6.82\,GHz.

To better test our prediction for the contribution from synchrotron
emission in our method, we use 1.4 GHz data, where synchrotron
emission is expected to be more significant.  We derive electron
temperatures using the VLA Galactic plane survey data set, which
covers the Galactic latitude range of the \citet{Balser2015} sample.
This comparison helped us to assess our derivation of electron
temperatures in sources in the southern hemisphere, where we used the
SGPS data set at 1.4\,GHz.  We convolved the 60\arcsec\ resolution
VGPS survey to 87\arcsec\ to match the angular resolution of the GBT
data set.  In the right panel of Figure~\ref{fig:Balser_comparison} we
show a comparison between the electron temperatures derived using our
method against those derived from the \citet{Balser2015} data set.  We
find that electron temperatures have a similar range with scatter
around the one-to-one correlation. In this case, the difference
between electron temperatures derived using our method and those from
\citet{Balser2015} is on average 33\%. As before, these differences
can arise from uncertainties in the relative calibrations between the
VGPS and GBT data sets and/or from the separation between free-free
and synchrotron emission, which is on average 38\% of the continuum
emission at 1.4\,GHz in this sample.

In summary, a comparison between our method to derive electron
temperatures using 1.4\,GHz and 6.82\,GHz observations shows that
additional uncertainties between 20\% to 30\% can arise due to
uncertainties in the calibration of several continuum bands and in our
correction for the contribution of synchrotron emission. Note that
these uncertainties are reduced by using a combination of several
radio continuum data sets in the derivation of electron temperatures
as discussed in Section~\ref{sec:electr-temp-determ}.

\bibliographystyle{aasjournal}

\bibliography{papers3}

\clearpage

\end{document}